\documentclass[twocolumn,showpacs,floatfix,superscriptaddress,nofootinbib,aps,prc,eqsecnum]{revtex4}
\usepackage[dvips]{graphicx}
\usepackage{latexsym,amssymb,amsmath,epsfig,bm,times,psfrag,subfigure}
\usepackage{color}
\usepackage[bookmarksnumbered,bookmarksopen,colorlinks,citecolor=blue,linkcolor=blue]{hyperref}

\renewcommand\a{\alpha}

\renewcommand\r{\rho}

\newcommand\e{\epsilon}
\newcommand\g{\gamma}

\newcommand\p{\pi}

\newcommand\f{\phi}

\newcommand\J{\Psi}

\newcommand\ls{\left[}
\newcommand\rs{\right]}
\newcommand\lc{\left\{}
\newcommand\rc{\right\}}
\newcommand{\lan}{\langle}
\newcommand{\ran}{\rangle}

\newcommand{\br}{{\mathbf r}}

\newcommand{\bv}{{\bf v}}

\newcommand{\bB}{{\bf B}}
\newcommand{\bE}{{\bf E}}

\renewcommand{\part}{{\rm part}}

\renewcommand{\vec}{\boldsymbol}
\newcommand{\be}{\begin{equation}}
\newcommand{\ee}{\end{equation}}
\newcommand{\bear}{\begin{eqnarray}}
\newcommand{\eear}{\end{eqnarray}}
\newcommand{\ba}{\begin{array}}
\newcommand{\ea}{\end{array}}

\begin{document}

\title{Azimuthally fluctuating magnetic field and its impacts on observables in heavy-ion collisions}

\author{John Bloczynski}
\affiliation{Physics Department and Center for Exploration of Energy and Matter,
Indiana University, 2401 N Milo B. Sampson Lane, Bloomington, IN 47408, USA.}

\author{Xu-Guang Huang}
\affiliation{Physics Department and Center for Exploration of Energy and Matter,
Indiana University, 2401 N Milo B. Sampson Lane, Bloomington, IN 47408, USA.}

\author{Xilin Zhang}
\affiliation{Physics Department and Center for Exploration of Energy and Matter,
Indiana University, 2401 N Milo B. Sampson Lane, Bloomington, IN 47408, USA.}

\author{Jinfeng Liao}
\affiliation{Physics Department and Center for Exploration of Energy and Matter,
Indiana University, 2401 N Milo B. Sampson Lane, Bloomington, IN 47408, USA.}
\affiliation{RIKEN BNL Research Center, Bldg. 510A, Brookhaven National Laboratory, Upton, NY 11973, USA.}


\begin{abstract}
The heavy-ion collisions can produce extremely strong transient magnetic and electric fields.   We study the azimuthal fluctuation of
these fields and their correlations with the also fluctuating matter geometry (characterized by the participant plane
harmonics) using event-by-event simulations. A sizable suppression of the angular correlations between the magnetic field and
the $2$nd and $4$th harmonic participant planes  is found in very central and very peripheral collisions,
while the magnitudes of these correlations peak around impact parameter $b\sim8-10\rm fm$ for RHIC collisions.
This can lead to notable impacts on a number of observables related to various magnetic field induced effects, and our finding suggests
that the optimal event class for measuring them  should be that corresponding to $b\sim8-10$ fm.
\end{abstract}
\pacs{12.38.Mh, 25.75.-q, 25.75.Ag, 24.10.Nz}
\maketitle

\section {Introduction}\label{intro}
Ultra-relativistic heavy-ion collisions create not only a domain of extremely high energy density where a new state of
matter --- the deconfined quark-gluon plasma (QGP) may form, but also extremely
strong (electro)magnetic fields due to the relativistic motion of the colliding heavy ions carrying large positive electric charge~\citep{Rafelski:1975rf}. Previous computations showed that the magnetic fields generated in Au + Au collision at RHIC ($\sqrt{s}=200\rm GeV$) can reach about
$eB\sim m_\p^2\sim 10^{18}$ G~\citep{arXiv:0711.0950,arXiv:0907.1396,arXiv:1003.2436,arXiv:1103.4239,arXiv:1107.3192,arXiv:1111.1949,arXiv:1201.5108}, which is $10^{13}$ times larger than the strongest man-made steady magnetic field in the laboratory. The magnetic field generated at LHC
energy can be an order of magnitude larger than that at RHIC~\citep{arXiv:0907.1396,arXiv:1201.5108}, according to a simple scaling law recently found in ~\citep{arXiv:1201.5108} for the event-averaged magnetic field: $\lan eB_y\ran\propto Zb\sqrt{s}$ for $b\lesssim2 R_A$ , where
$Z$ is the charge number of the ions, $b$ is the impact parameter, $R_A$ is the radius of nucleus, and
the $y$-axis is perpendicular to the reaction plane. Thus, heavy-ion collisions provide a unique terrestrial environment with ultra-strong magnetic fields.

There have been very strong interests and intensive efforts recently in studying various possible physical effects induced by the presence of strong magnetic field. Receiving particular enthusiasm is the set of ideas to look for experimental manifestation of QCD effects stemming from topology and anomaly and aided by the external magnetic field. These include, e.g., the so-called Chiral Magnetic Effect (CME)~\citep{Kharzeev:2004ey,Kharzeev:2007tn,arXiv:0711.0950,Fukushima:2008xe} in which a nonzero axial charge density in the matter (presumably from topological transitions via sphalerons) together with the magnetic field $\bB$ will induce a dipole charge separation along the
$\bB$ direction. A lot of works have been done  in the past few years to experimentally measure this effect by analyzing the charged particle correlations~\citep{Voloshin:2004vk,Abelev:2009ac,Selyuzhenkov:2011xq} and to understand the interpretation and background effects related to these measurements~\citep{Schlichting:2010qia,Pratt:2010zn,Bzdak:2009fc,Bzdak:2010fd,Liao:2010nv,Wang:2009kd}, as recently reviewed in Ref.~\citep{Bzdak:2012ia}. There is also the so-called Chiral Separation Effect (CSE)~\citep{Son:2004tq}, from its interplay with the CME there arises a gapless collective excitation called the Chiral Magnetic Wave (CMW)~\citep{Kharzeev:2010gd}. An observable effect of CMW was proposed in ~\citep{Burnier:2011bf}:   the CMW with the presence of nonzero vector charge density and the magnetic field transports  the charges in QGP towards an electric quadrupole distribution with more positive charges near the poles of  the produced fireball (pointing outside of the reaction plane) while more negative charges near the equator (in the reaction plane), and this leads to a measurable splitting of negative and positive pions' elliptic flow.  Recently STAR
collaboration reported the first measurement of the charged pion flow spitting versus the charge asymmetry which appears in agreement with the predictions from CMW~\citep{Burnier:2011bf,Burnier:2012ae}.
Yet one more interesting effect is possible soft photon production through the QCD conformal anomaly in the external magnetic field as suggested in ~\citep{Basar:2012bp}. (The CME could also possibly cause anisotropic soft photon production, see~\cite{Fukushima:2012fg}.) In this mechanism the photons are emitted perpendicular to the $\bB$ direction and there is an appreciable azimuthal anisotropy of the emitted photons which might partially account for the unusually large $v_2$ of direct photons reported by PHENIX collaboration~\citep{Adare:2011zr}.
In addition to the above anomaly phenomena in magnetic field, there are also discussions of other novel effects in  strong magnetic fields~\citep{arXiv:1008.1055,arXiv:1102.3819,arXiv:1108.4394,arXiv:1108.0602}, e.g., the spontaneous electromagnetic
superconductivity of QCD vacuum, the possible enhancement of elliptic
flow of charged particles, possible anisotropic electromagnetic radiation from the QGP, the energy loss due to the
synchrotron radiation of quarks, and the emergence of anisotropic viscosities.

In most of the phenomenological studies of these effects, the magnetic field $\bB$ in heavy-ion collisions represents the largest source of uncertainty in their quantitative calculations, and a precise knowledge of the magnetic field is urgently needed. This is particularly so in the context of the realization and intensive investigations in the last two or three years that there are very strong fluctuations in the initial conditions of heavy-ion collisions. Such fluctuations have been shown to lead to strong observable effects both in the bulk collective expansions (as ``harmonic flows'')~\citep{harmonic_flow}  and in the anisotropy of penetrating hard probe (as ``harmonic tomography'')~\citep{harmonic_tomography}. Since the magnetic field directly relies on the initial distributions of  protons in both nuclei,  a realistic computation has to take into account such strong initial fluctuations. A first step has been made in \citep{arXiv:1111.1949,arXiv:1201.5108} to compute the magnitude of $\bB$ with event-by-event fluctuations, which was indeed found to be remarkably modified from the naive ``optical'' estimates.

In this paper, we focus on the strong fluctuations in the azimuthal orientation of the magnetic field $\bB$, and particularly investigate its angular correlation with the  underlying matter geometry (specified by participant planes)  bearing concurrent fluctuations on an event-by-event basis.  This correlation is a very important  link between the experimental measurements and any of the  magnetic field induced effects in heavy-ion collisions. Most (if not all) previous studies rely on the assumption that the magnetic field is pointing in the out-of-plane direction which would be true without fluctuations. In the real world, however, both the $\bB$ and the ``planes'' (event planes or participant planes) bear strong fluctuations and as we will show their orientations are never fully ``locked'' and only strongly correlated in some circumstances. We will study these correlations in great details and examine the consequences for a number of observables related to some aforementioned effects.

The rest of this paper is organized as follows. In Sec.~\ref{event} we show how the initial fluctuations modify the magnetic and electric fields generation
in heavy-ion collisions. We study the azimuthal fluctuation of the magnetic field in Sec.~\ref{azimu}. We discuss the physical
implications of these results in Sec.~\ref{impac}. For completeness we also show results for the electric field in Sec.~\ref{elect}. In sec.~\ref{qnc} a new class of charge dependent measurements partly motivated by our results will be discussed. Finally we summarize in Sec.~\ref{discu}. The natural unit $\hbar=c=k_B=1$ will be used throughout this article.

\section {Event-by-event calculation of the electromagnetic field}\label{event}
The aim of this section is to give an event-by-event calculation of the electromagnetic fields in
heavy-ion collisions. We focus on the fields
at the initial time $t=0$, that is, the moment when the two colliding nuclei overlap completely.
Our starting point is the Lorentz boosted Coulomb formulas which are equivalent to the
Li\'enard-Wiechert potentials for constantly moving charges:
\begin{eqnarray}
\label{LWB}
e\bB(t,\br)&=&\frac{e^2}{4\p}\sum_n Z_n({\bf R}_n)\frac{1-v_n^2}{[R_n^2-({\bf R}_n\times\bv_n)^2]^{3/2}}{\bf v}_n\times{\bf R}_n,\nonumber \\
\label{LWE}
e\bE(t,\br)&=&\frac{e^2}{4\p}\sum_n Z_n({\bf R}_n)\frac{1-v_n^2}{[R_n^2-({\bf R}_n\times\bv_n)^2]^{3/2}}{\bf R}_n,
\end{eqnarray}
where ${\bf R}_n=\br-\br_n(t)$ is the relative
position of the field point $\br$ to the $n$th proton at time $t$, $\br_n(t)$, and $\bv_n$ is the velocity of the $n$th proton.
The summations run over all protons in the projectile and target nuclei. Equations (\ref{LWE})contain
singularities at $R_n=0$ if we treat protons as point charges. In practical
calculation, to avoid such singularities we treat protons as uniformly charged spheres with radius $R_p$. The charge number factor $Z_n(\bf{R}_n)$ in Eqs.~(\ref{LWE}) is introduced to encode this aspect: when the field point locates outside the $n$th proton (in the rest frame of the proton) $Z_n=1$, otherwise $Z_n<1$ depends on ${\bf R}_n$. The in-medium charge radius $R_p$ of proton is unknown (the most recent measurement of the rms charge radius of proton gives $R_p=0.84184(67)$ fm in vacuum~\cite{Pohl:2010zza}), we choose $R_p=0.7$ fm in our numerical simulations. We checked that varying $R_p$ from $0.6$ fm to 0.9 fm will shift the numerical results within $15\%$ but no qualitative conclusion is altered.
The nucleons in one nucleus move at constant velocity along the beam direction (we choose it as $z$-direction), with the nucleons
in another nucleus moving oppositely at the same speed. The energy for each nucleon is set to be $\sqrt{s}/2$ in the
center-of-mass frame, therefore the value
of the velocity of each nucleon is given by $v_n^2=1-(2m_N/\sqrt{s})^2$, where $m_N$ is the mass of the nucleon. We set
the $x$-axis along the impact parameter vector so that the reaction plane is the $x$-$z$ plane. Finally,
the positions of nucleons in the rest frame of a nucleus are sampled according to the Woods-Saxon distribution.

\begin{figure}
\begin{center}
\includegraphics[width=5.5cm]{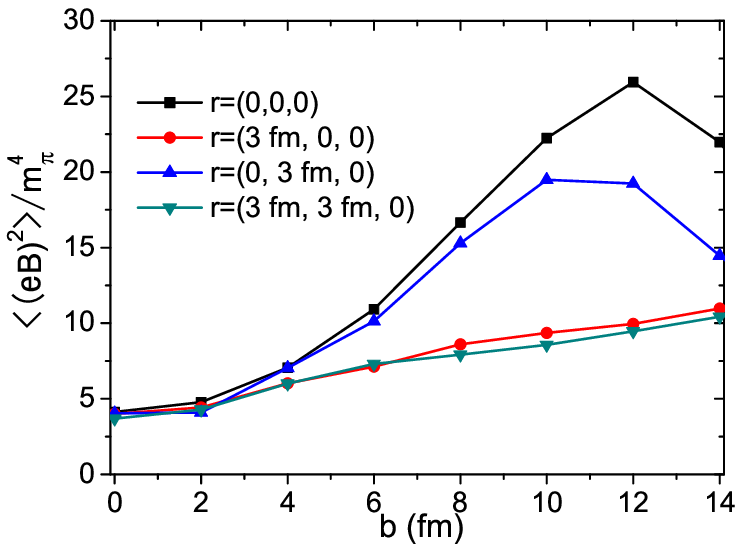}
\includegraphics[width=5.5cm]{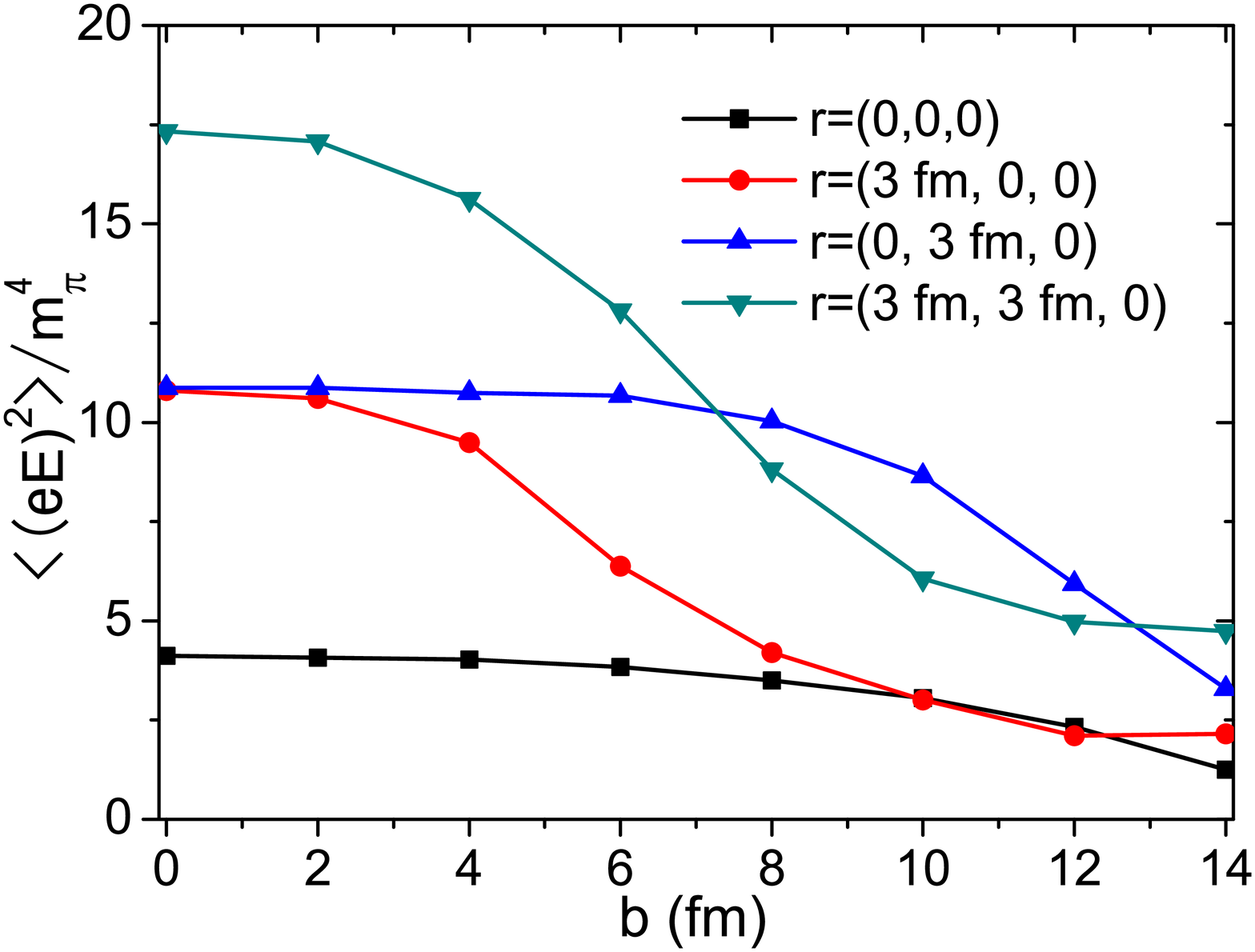}
\caption{(Color online) The event-averaged $(eB)^2$ and $(eE)^2$ (in unit of $m^4_\pi$) at $t=0$ and four different points on the
transverse plane
as functions of the impact parameter $b$.}
\label{field2}
\end{center}
\vspace{-0.7cm}
\end{figure}

\begin{figure*}
\begin{center}
\includegraphics[width=4.4cm]{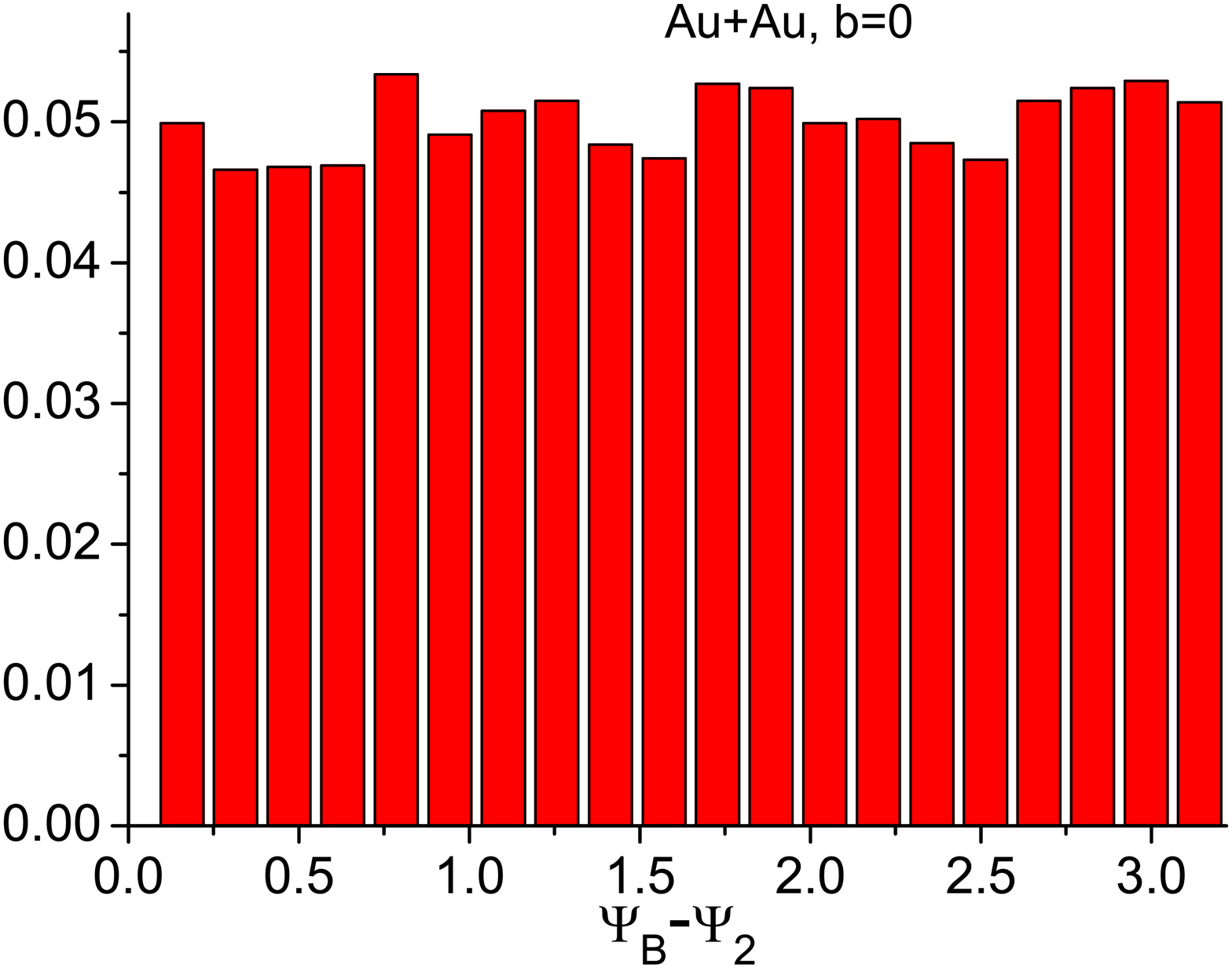}
\includegraphics[width=4.4cm]{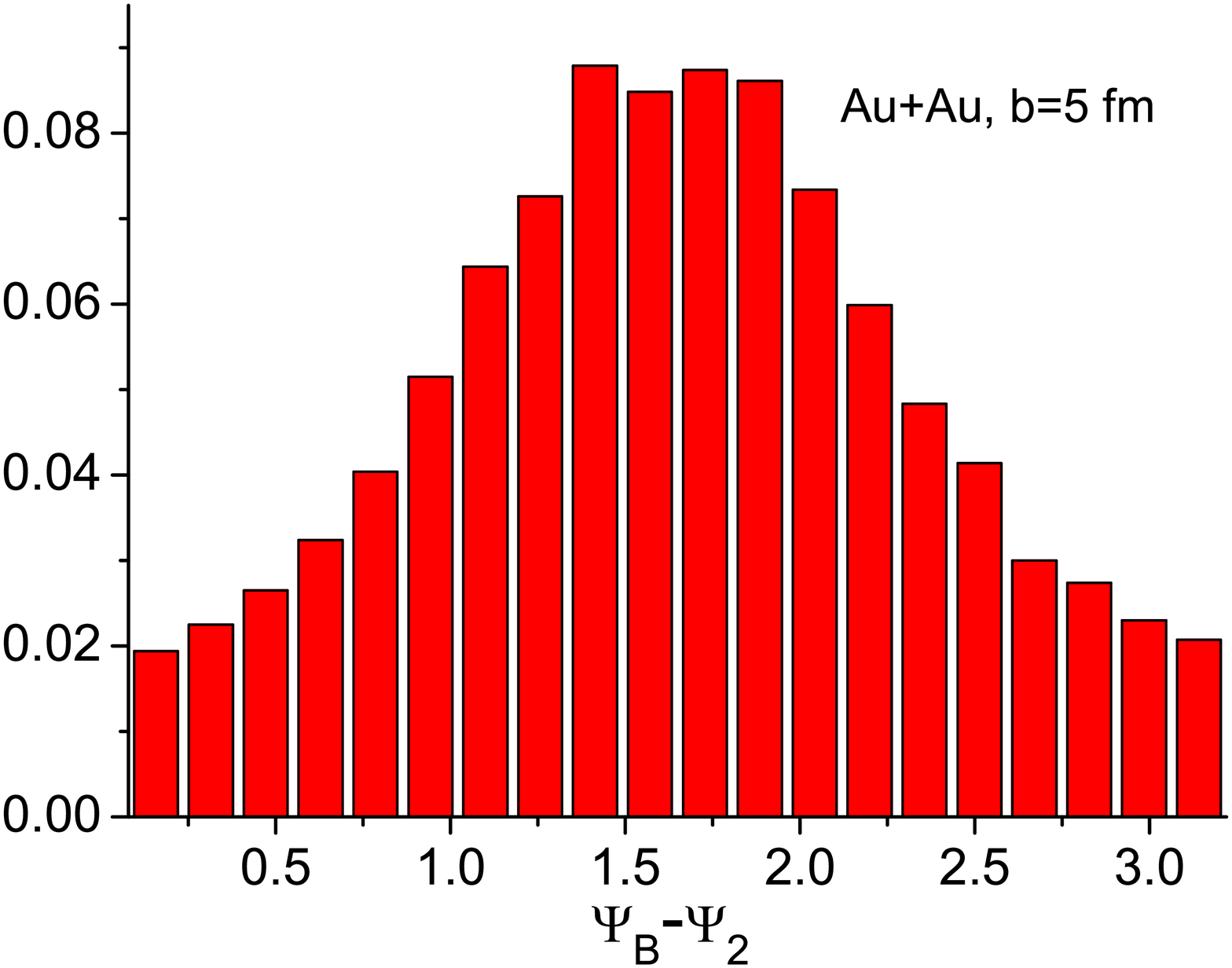}
\includegraphics[width=4.4cm]{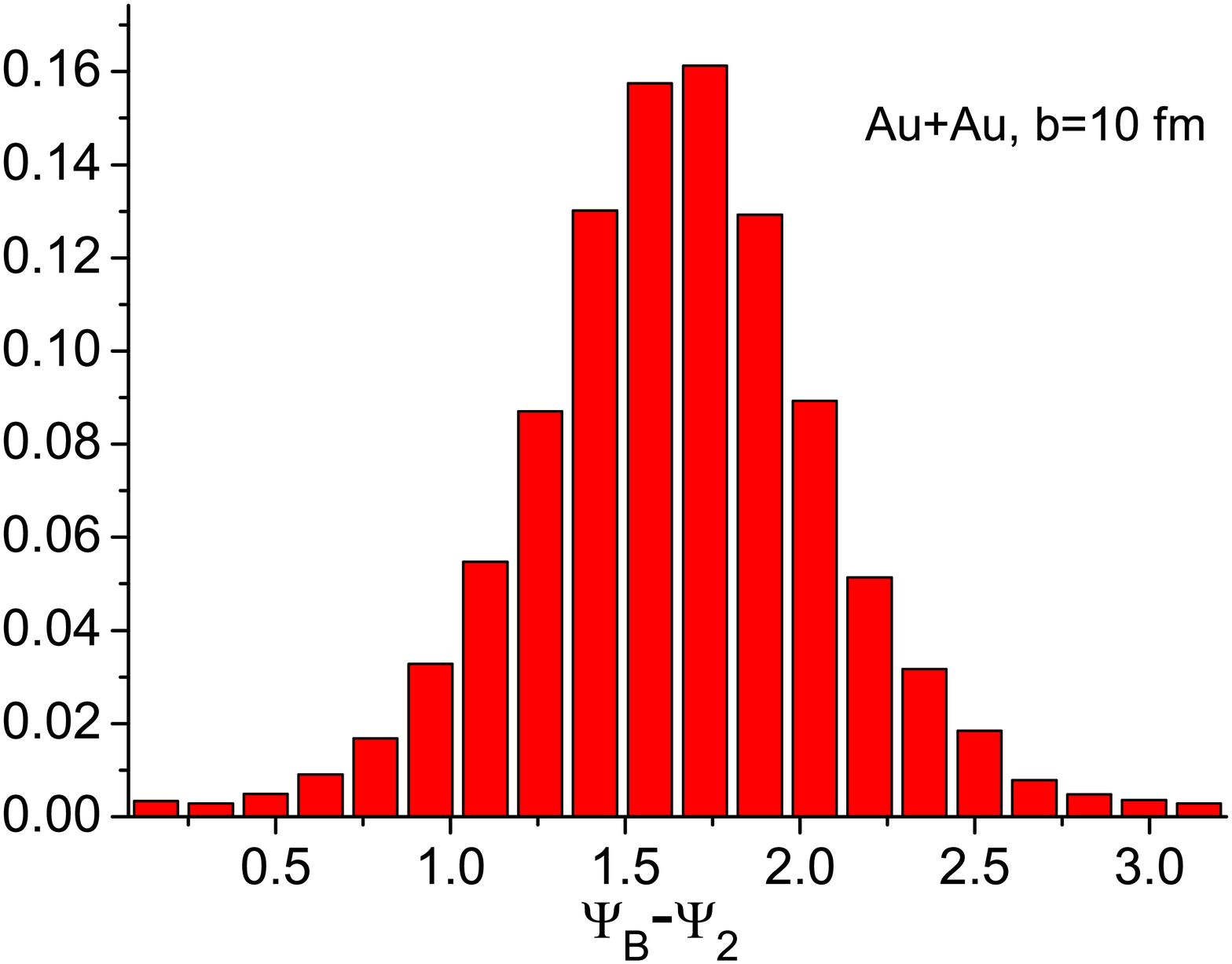}
\includegraphics[width=4.4cm]{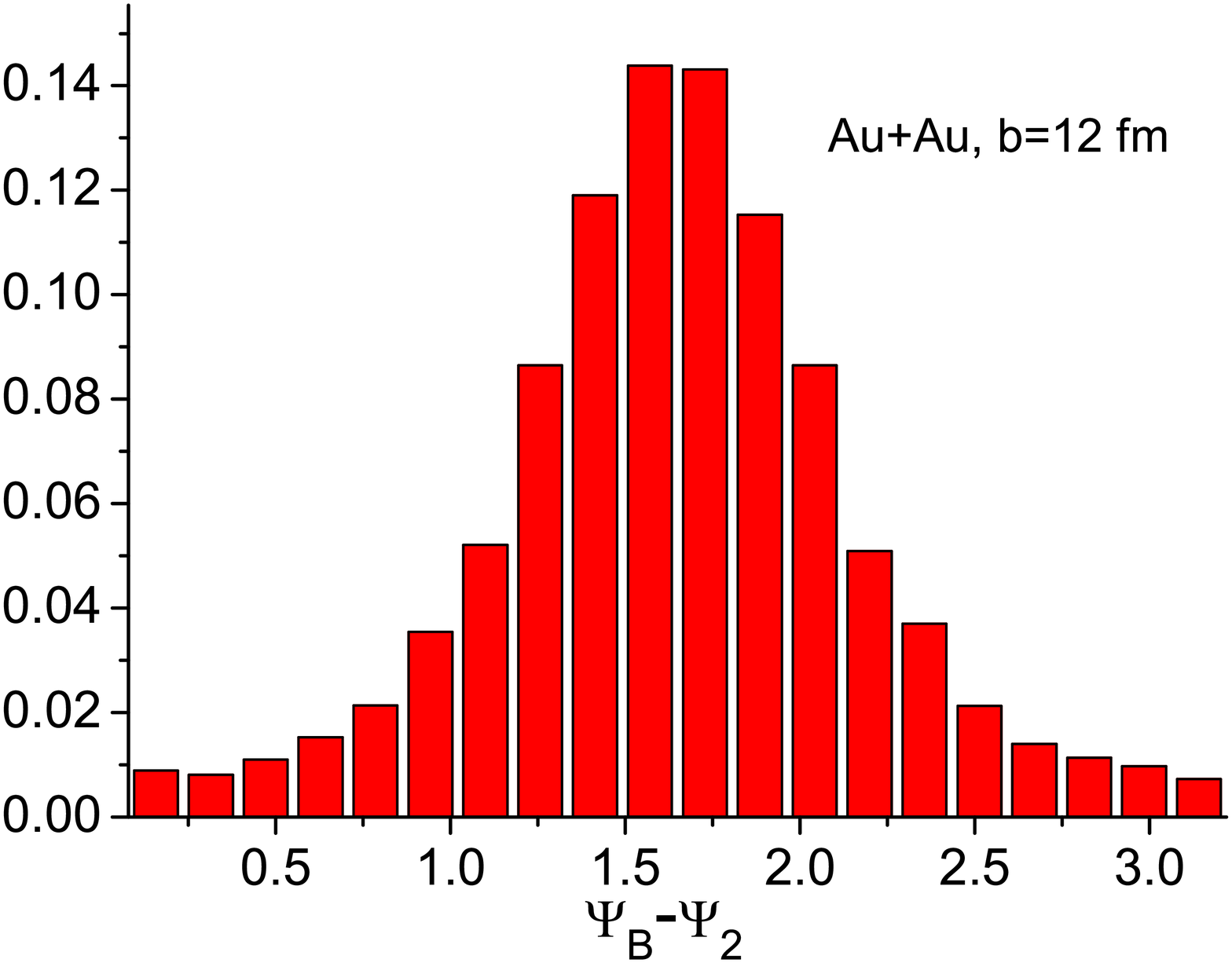}
\caption{(Color online) The event-by-event histograms of $\J_\bB-\J_2$ at impact parameters $b=0, 5, 10, 12$ fm for Au + Au collision at RHIC energy. Here $\J_\bB$ is the azimuthal direction of $\bB$ field (at $t=0$ and $\br=(0,0,0)$) and
$\J_2$ is the second harmonic participant plane.}
\label{his}
\end{center}
\vspace{-0.2cm}
\end{figure*}

\begin{figure*}
\begin{center}
\includegraphics[width=4.4cm]{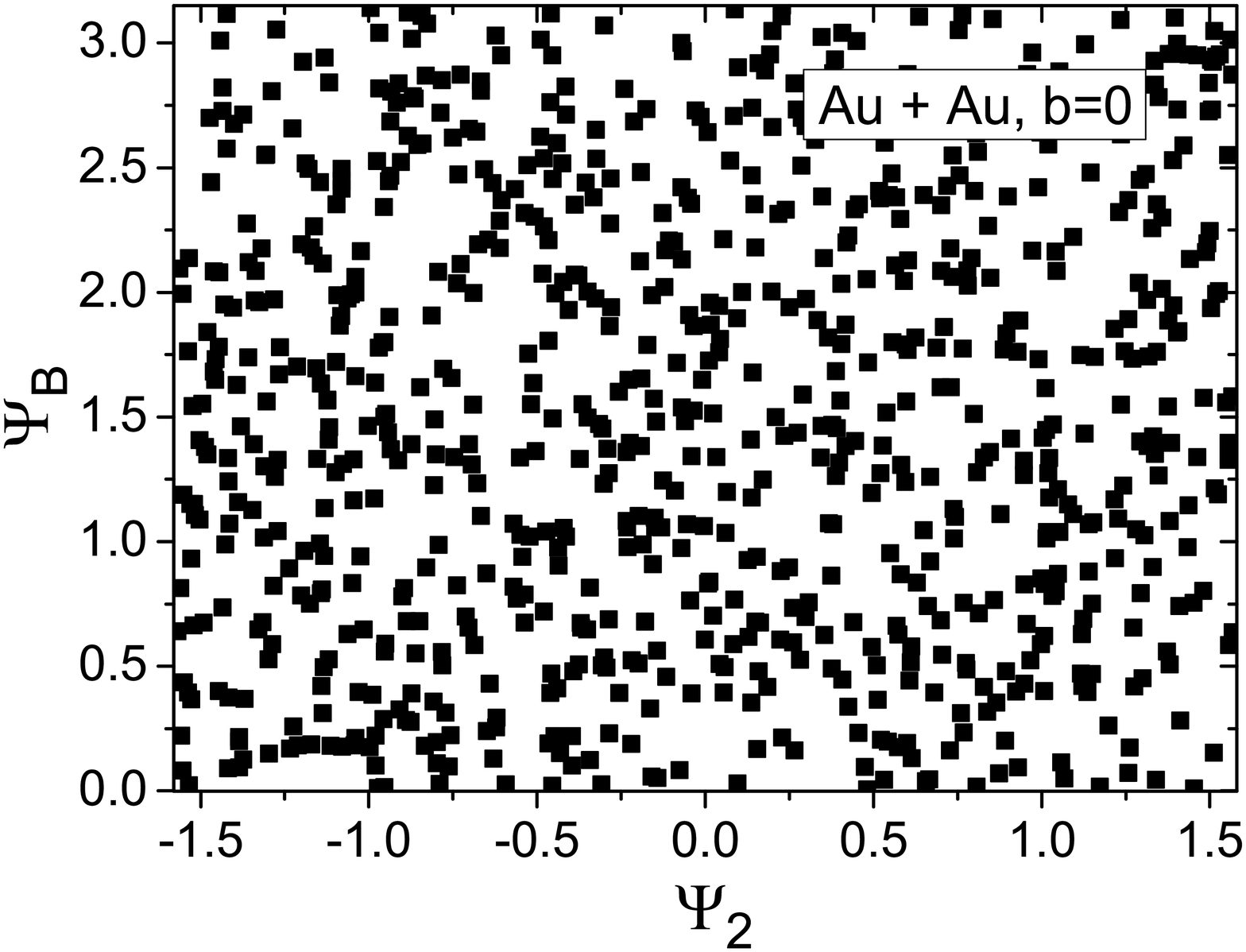}
\includegraphics[width=4.4cm]{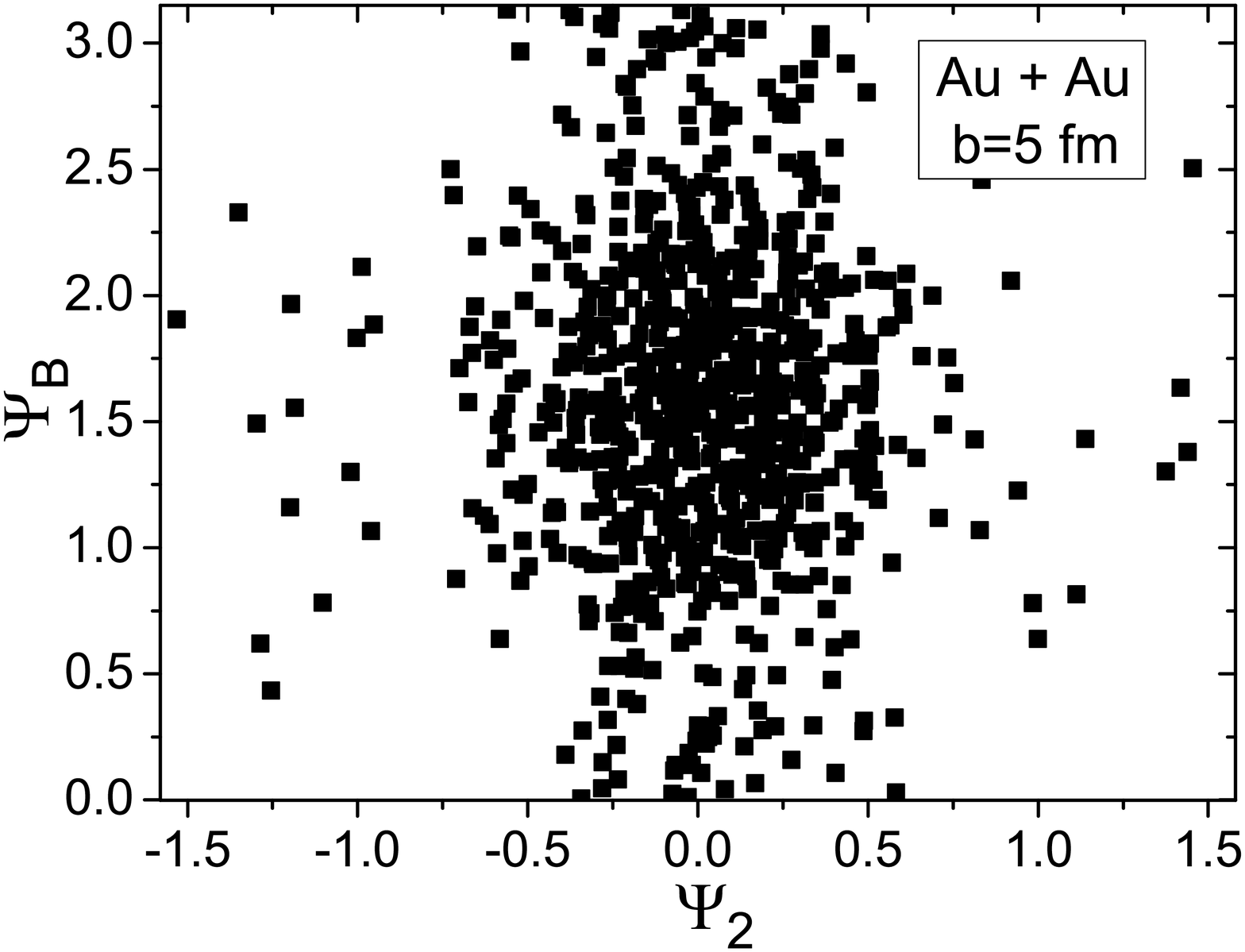}
\includegraphics[width=4.4cm]{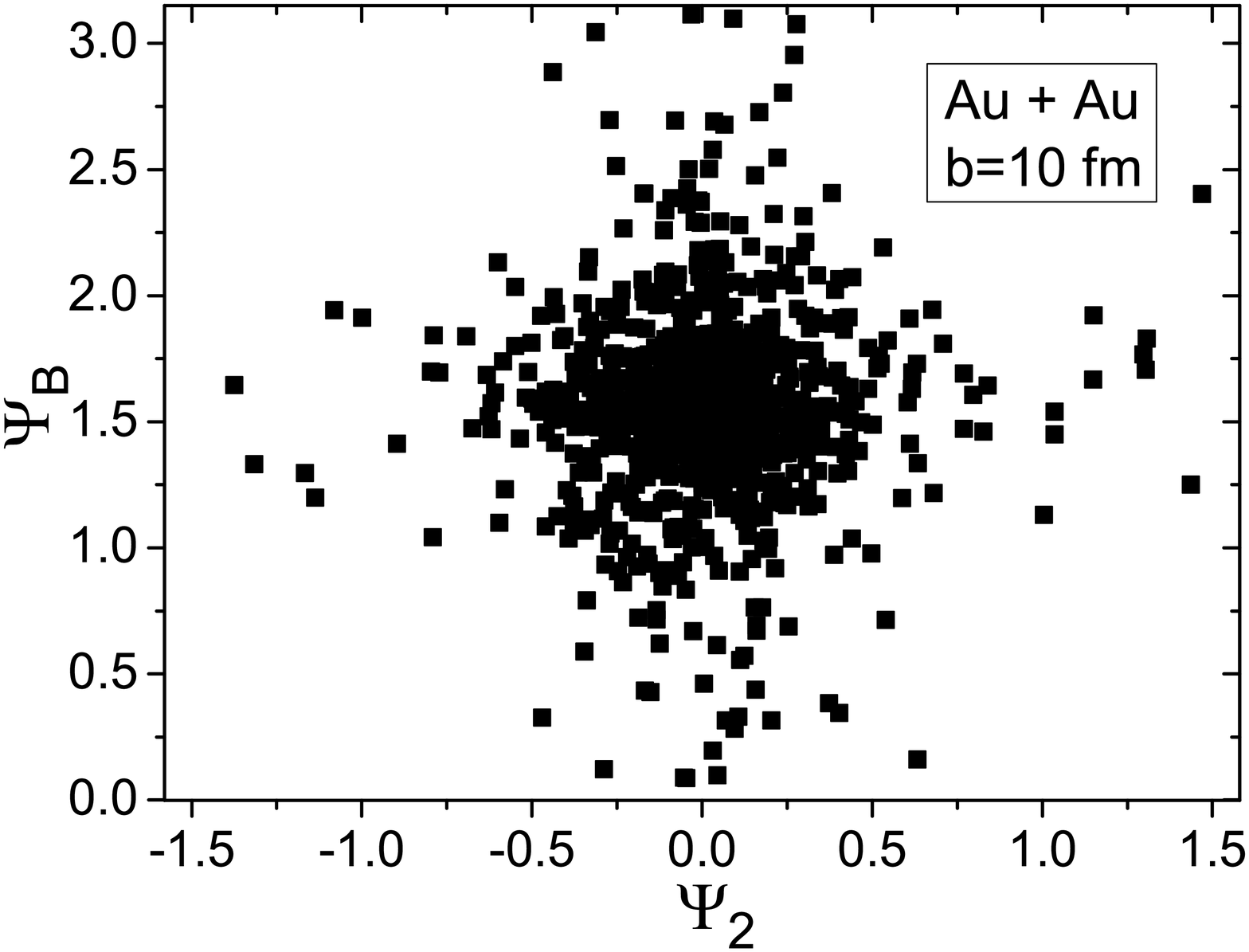}
\includegraphics[width=4.4cm]{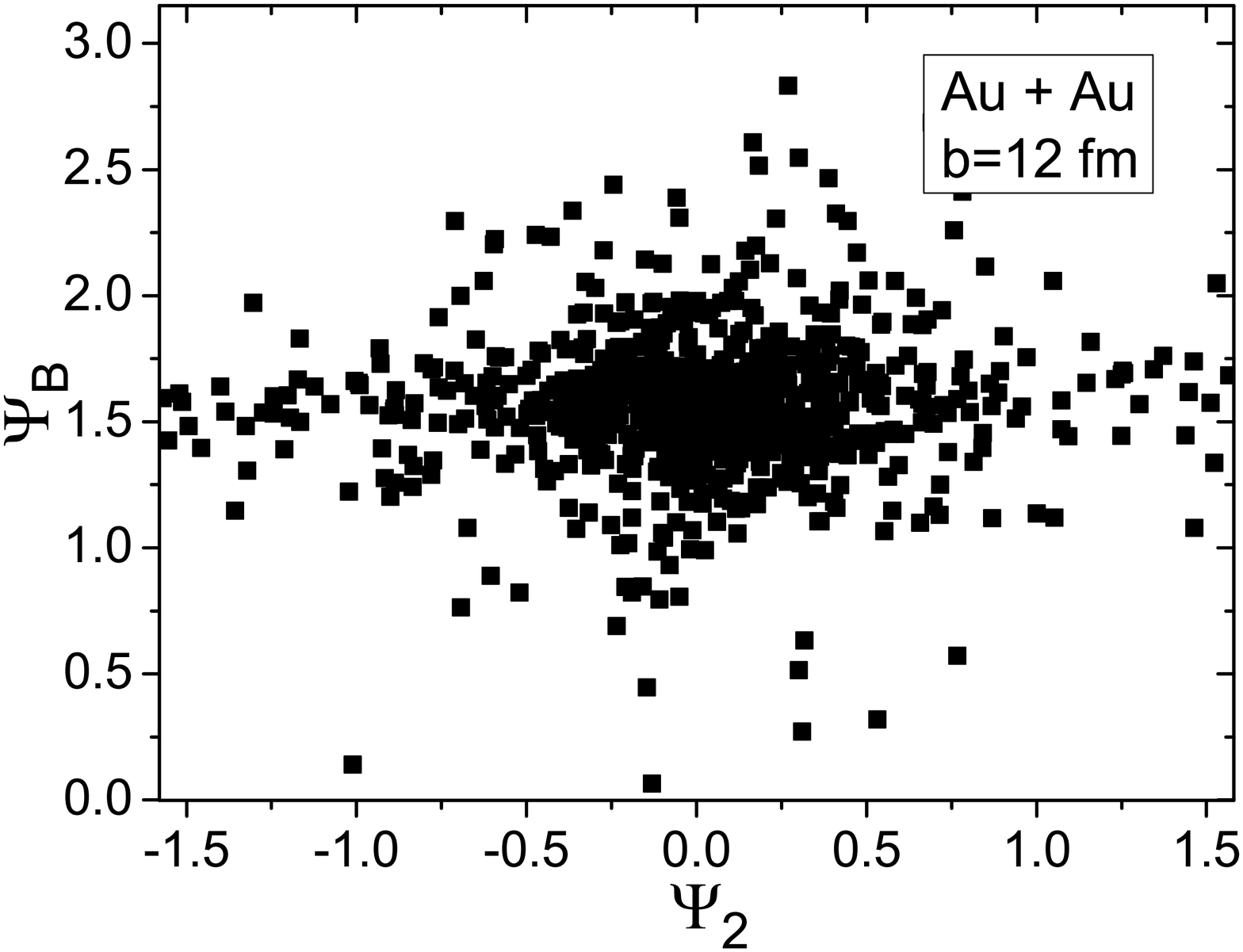}
\caption{The scatter plots on $\J_B$-$\J_2$ plane  at impact parameters $b=0, 5, 10, 12$ fm for Au + Au collision at RHIC energy. Here $\J_\bB$ is the azimuthal direction of $\bB$ field (at $t=0$ and $\br=(0,0,0)$) and
$\J_2$ is the second harmonic participant plane.}
\label{scat}
\end{center}
\vspace{-0.7cm}
\end{figure*}

In this computation we have obtained in each event all the components of both the $\bB$ and $\bE$ fields at  four different points, $\br=(0,0,0), \br=(3\, {\rm fm}, 0,0 ), \br=(0, 3\, {\rm fm}, 0)$, and $\br=(3\, {\rm fm}, 3\, {\rm fm}, 0)$, on the transverse plane for a wide range of impact parameter $b$ in Au + Au collision
at RHIC energy $\sqrt{s}=200$ GeV.  We have found that all the  $z$ components are negligibly small, while the transverse components are strong, with their centrality trends in agreement with previous results in ~\citep{arXiv:1111.1949,arXiv:1201.5108}. The absolute magnitudes of these fields are  somewhat smaller than those reported in  ~\citep{arXiv:1111.1949,arXiv:1201.5108}, due to the important difference that we treat each proton as a charged sphere with radius $R_p$ rather than a point charge, which is both more physical and mathematically less of singularity problem. As also noted in  ~\citep{arXiv:1111.1949,arXiv:1201.5108} and confirmed in our computation, although the $x$-component of the magnetic field as well as the $x$- and $y$-components of the electric field vanish after averaging over many events, their magnitudes in each event can be huge and comparable to the $y$-component of the magnetic field due to the fluctuations. This already implies that  both the magnitude and the direction of the electromagnetic field (albeit in the unmeasurable $x$-$y$ plane) fluctuate strongly. In Fig.~\ref{field2} we show $\lan(eB)^2\ran$ and $\lan(eE)^2\ran$ as functions of $b$ at different points: clearly the electromagnetic fields bear considerable inhomogeneity on the transverse plane. We also notice that the electric field can be very strong particularly for more central collisions. The $\lan(eB)^2\ran$ is particularly interesting because the signal strengths of several magnetic-field-induced effects are  proportional to $(eB)^2$, as we will discuss more in Sec.~\ref{impac}.


\section {Azimuthal correlations between magnetic field and matter geometry}\label{azimu}
As already mentioned before, on the event-by-event basis the electromagnetic field fluctuates strongly both in magnitude
and in azimuthal direction. The direction of the $\bB$-field is very important as  the $\bB$-field induced effects inherit this information and occur either along (e.g. CME, CMW) or perpendicular (e.g. photon emission) to $\bB$.

Even more important from the measurement perspective is the question of  the $\bB$ orientation with respect to a frame that could be identified experimentally rather than to the ideal world reaction plane. In reality what could be determined is the final-state hadrons' distribution geometry in momentum space, in particular the second harmonic $v_2$ event plane (EP) which is relatively more tightly correlated with the initial matter distribution geometry that can be specified primarily by the second harmonic $\epsilon_2$ participant plane (PP). Therefore the really useful information is the $\bB$ orientation with respect to the matter geometry per event. Here we make a first detailed examination of such a kind,  studying the azimuthal correlation between the magnetic field and the participant planes.

In order to do that, we determine the participant planes for various harmonics from the Monte Carlo Glauber simulations of the initial condition and analyze the angular correlations between
the $\bB$ and the participant plane orientations from the same event. The  $n$th harmonic participant plane angle $\J_n$ and eccentricity $\e_n$ are calculated from participant density $\r(\br)$ as in the literature (e.g.~\citep{harmonic_flow}):
$\e_1e^{i\J_1} =-(\int d^2\br_\perp\r(\br_\perp) r_\perp^3 e^{i\f})/({\int d^2\br_\perp\r(\br_\perp) r_\perp^3}),$ and
$\e_n e^{in\J_n}=-({\int d^2\br_\perp\r(\br_\perp) r_\perp^n e^{in\f})/(\int d^2\br_\perp\r(\br_\perp) r_\perp^n})$ for $n>1$.
In this Section we focus on the 2nd harmonic participant plane $\J_2$ as it is the most prominent anisotropy from both geometry and fluctuations. Correlations of $\bB$ with  other harmonics will be discussed in Sec.~\ref{impac}.
By determining $\bB$ and $\J_2$ in each event we can examine the distribution of their relative angle over many events:  in Fig.~\ref{his} we plot the event-by-event histograms of $\J_\bB-\J_2$ at $b=0,5,10$ and $12$ fm for Au + Au collisions at $\sqrt{s}=200$ GeV where $\J_\bB$ is the azimuthal direction of the magnetic field at $t=0$ and $\br=(0,0,0)$.
Strikingly at $b=0$ the histogram of $\J_\bB-\J_2$ is almost uniform indicating that $\J_\bB$ and $\J_2$ are basically uncorrelated. For $b=5, 10$ and $12$ fm, the histograms have Gaussian shapes peaking at $\p/2$  with the corresponding widths not small at all. The width first decreases when $b$ increased from $5$ to $10$ fm and then increases again toward $b=12$ fm. So although the $\bB$ field indeed points at $\pi/2$ with respect to $\J_2$ plane on average, it fluctuates significantly in each  event. The  correlation between $\J_\bB$ and $\J_2$ is the strongest in middle-centrality while weakens much in the most central and most peripheral collisions.

To further reveal the correlation pattern between $\J_\bB$ and $\J_2$
and to understand better the non-monotonous centrality trend of the widths of the
histograms in Fig.~\ref{his},  we show the scatter plots  from all events at given $b$ on
the $\J_\bB$-$\J_2$ plane in Fig.~\ref{scat} which visualize the 2D probability distribution density. Again  for $b=0$, the events are almost
uniformly distributed indicating negligible correlation between
$\J_\bB$ and $\J_2$. For $b=5, 10$, and $12$ fm, the event distributions evidently concentrate
around $(\J_\bB, \J_2)=(\p/2,0)$ indicating a correlation between the two. Going from $b=5$ to $10$ and to $12$ fm, the spread in $\J_\bB$ keeps shrinking while the spread in $\J_2$ clear grows with larger $b$. This is because for non-central collisions with increasing $b$, the $\bB$ is mostly from the spectators whose number increases and bears less fluctuations while the $\J_2$ is determined by participants whose number decreases and fluctuates more. This explains the non-monotonic trend of the widths in the histograms in Fig.~\ref{his}.
In short, we have found that the event-by-event fluctuations of the initial condition bring
azimuthal fluctuations in both $\J_\bB$ and $\J_2$, and the angular  correlation between them is smeared out significantly in the very central and very peripheral collisions while stays strong for middle-centrality collisions. This observation certainly influences the interpretation of observables related with $\bB$-induced effects, as will be discussed  in the next section.

\section {Impact on various observables}\label{impac}

We now discuss the impacts of the azimuthal fluctuations of the magnetic field with respect to matter geometry (the participant planes)
on a number of pertinent observables  recently   measured in heavy-ion collisions. We recap a few points already discussed:  the $\bB$-induced effects  are sensitive to the azimuthal direction of the $\bB$ field, which however is not experimentally known;  in the past when observables are proposed and interpreted for measuring certain $\bf B$-related effects, it is often assumed that the $\bf B$ direction is  perpendicular to the reaction plane; as already shown in last section, this assumption is  not true, and the azimuthal orientation between $\Psi_{\bf B}$ of $\bf B$ and  $\Psi_2$ fluctuates  with sizable spread in their relative angle $(\Psi_{\bf B}-\Psi_2)$. In what follows we evaluate the impacts of such fluctuations on the experimentally measured quantities.

First, let us consider the pair correlation $\gamma=\lan\cos(\phi_1+\phi_2-2\Psi_{2})\ran$ with $\phi_{1,2}$ the azimuthal angles of the particle 1 and 2 where the average is taken over events. The measurements of $\gamma$ are motivated by the search for the Chiral Magnetic Effect~\citep{arXiv:0711.0950}. (The $\gamma$ is actually extracted through three particle correlations $\lan\cos(\phi_1+\phi_2-2\phi_3)\ran$ divided by elliptic flow $v_2$ with the third particle serving as a ``projector'' toward the Event Plane~\cite{Voloshin:2004vk}. ) Now let us focus on the same-charge pairs, and suppose the CME indeed gives rise to the same-side azimuthal correlations along the $\bf B$-field direction for the same-charge pairs. The two-particle density then receives the following contribution (with $A_{++}$ the signal strength)
\begin{eqnarray}
f_{++} = A_{++} \cos(\phi_1-\Psi_{\bf B}) \, \cos(\phi_2-\Psi_{\bf B}).
\end{eqnarray}
This translates into the following form after re-defining angles $\bar{\phi}_i = \phi_i - \Psi_2$ and $\bar{\Psi}_{\bf B}=\Psi_{\bf B}- \Psi_2$:
\begin{eqnarray}
f_{++} &=& \frac{A_{++}}{2}\cos(\bar{\phi}_1- \bar{\phi}_2) + \frac{A_{++}}{2} [\cos(2\bar{\Psi}_{\bf B})]\, \cos(\bar{\phi}_1+\bar{\phi}_2)  \nonumber \\ && +  \frac{A_{++}}{2} [\sin(2\bar{\Psi}_{\bf B})]\, \sin(\bar{\phi}_1+\bar{\phi}_2).
\end{eqnarray}
We therefore see that the CME's contribution to the  $\gamma_{++}$:
\begin{eqnarray}
\gamma_{++} \sim  \frac{\lan A_{++}  \cos(2\bar{\Psi}_{\bf B})\ran}{2}.
\end{eqnarray}
If the $\bf B$-direction were to be always perfectly aligned with the out-of-plane, i.e. $\bar{\Psi}_{\bf B}=\pi/2$, then we simply have $\gamma_{++}\to  - \lan A_{++}\ran/2$. But the fluctuations in magnetic field as well as in matter geometry will blur the angular relation between the two and modify the signal by the factor $\sim \lan \cos(2\bar{\Psi}_{\bf B})\ran$.
Similarly, if one measures the charge separation with respect to higher harmonic participant plane,
e.g. the fourth harmonic plane $\J_4$,  the corresponding correlation will be $\lan\cos[2(\phi_1+\phi_2-2\Psi_{4})]\ran$ (as recently
proposed with the hope to disentangle the collective-flow and CME contributions to $\gamma$~\cite{Voloshin:2011mx}). The azimuthal fluctuation of $\bB$ field with respect to $\J_4$ will again contribute a modification factor $\sim\lan\cos[4(\Psi_{\bf B}-\J_4)]\ran$ in the above  correlation.

\begin{figure*}
\begin{center}
\includegraphics[width=4.4cm]{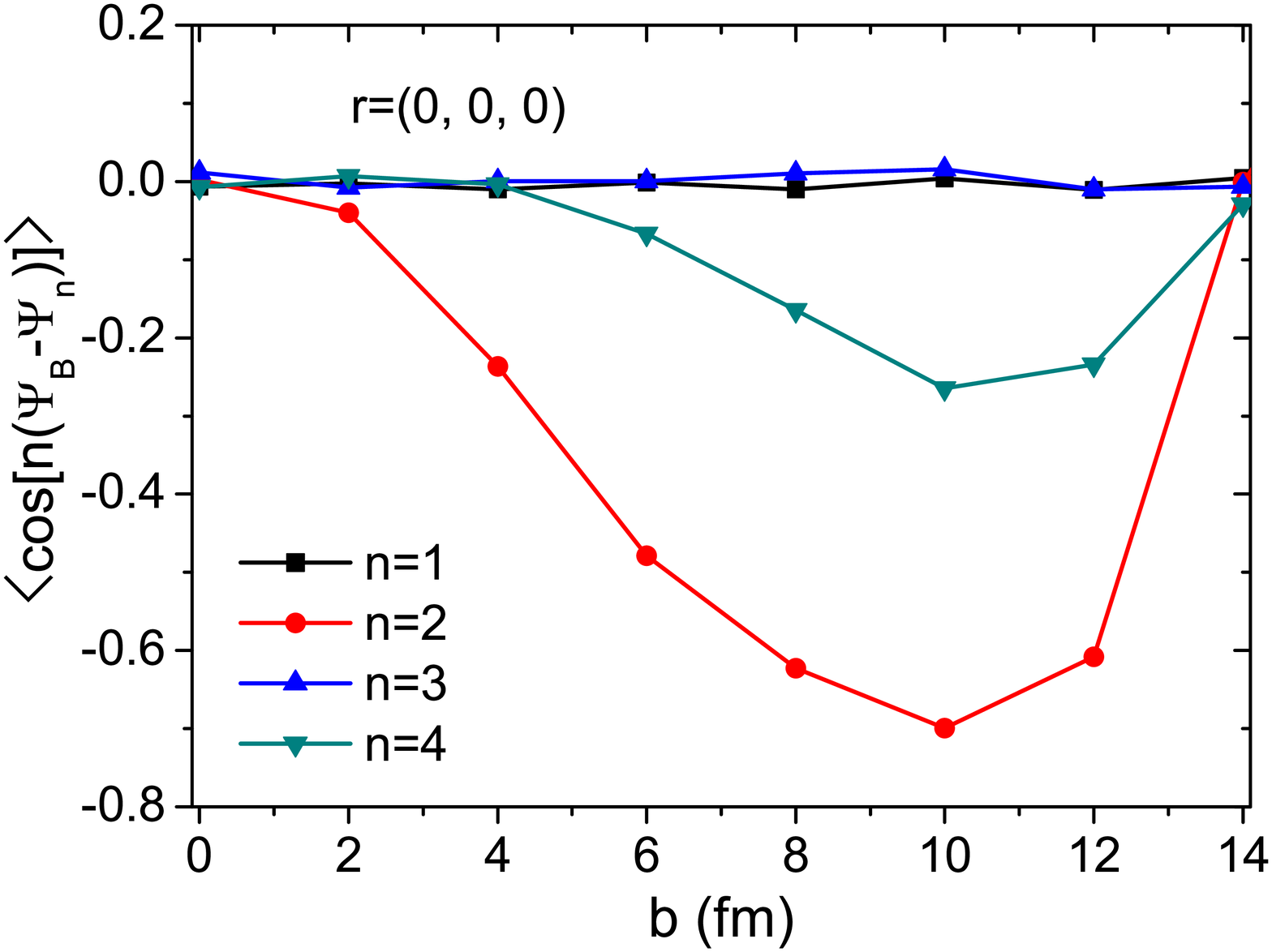}
\includegraphics[width=4.4cm]{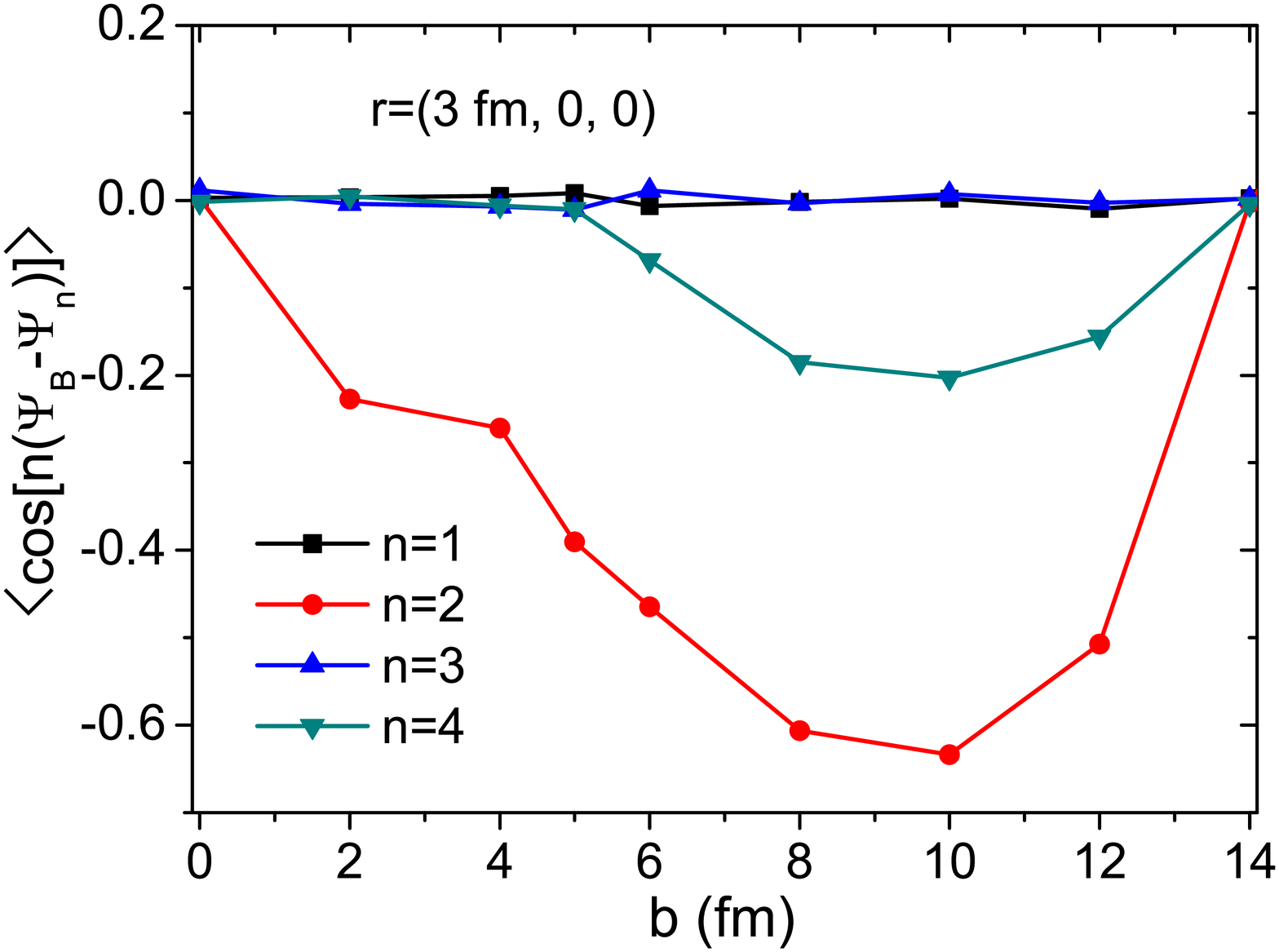}
\includegraphics[width=4.4cm]{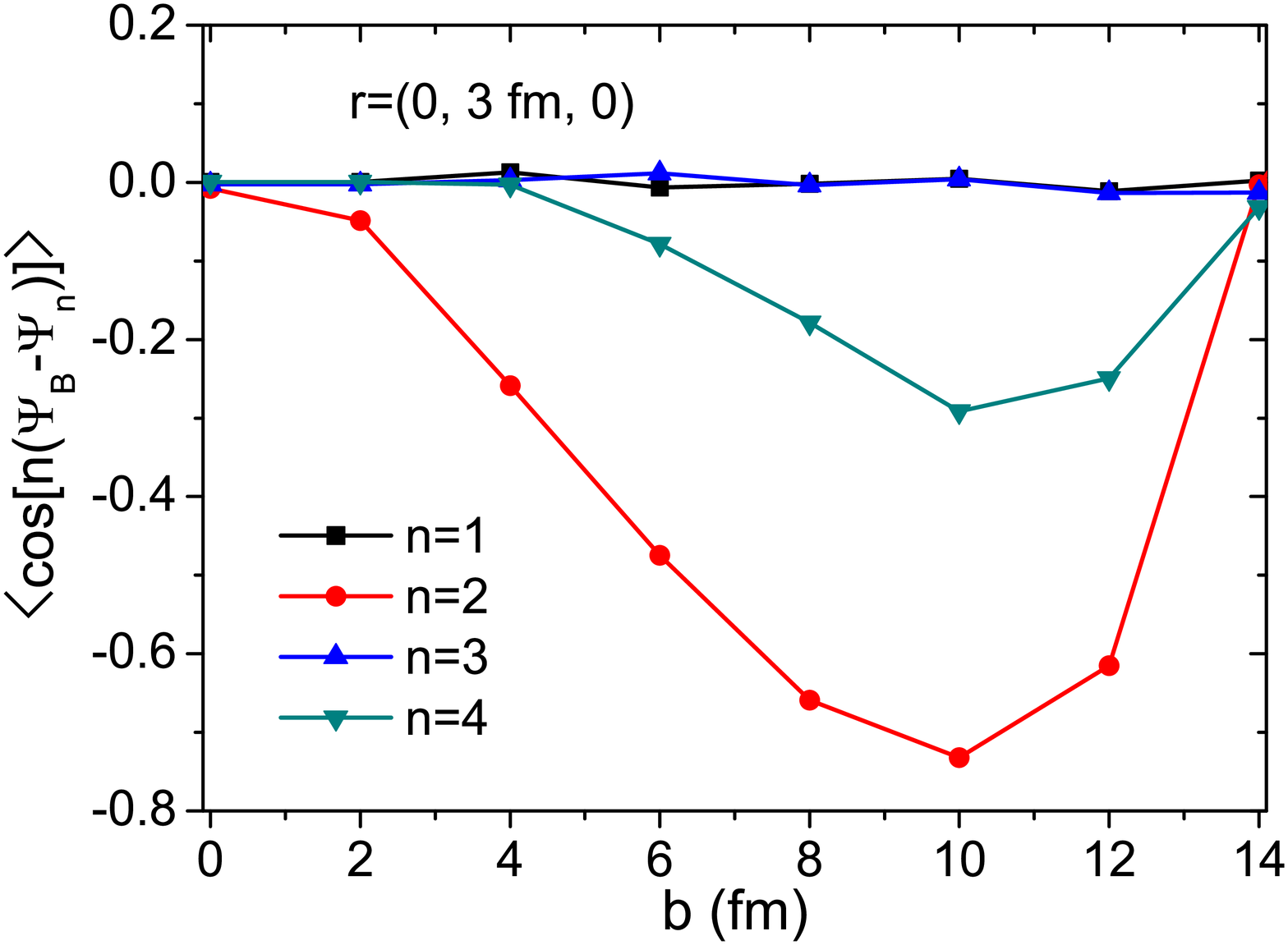}
\includegraphics[width=4.4cm]{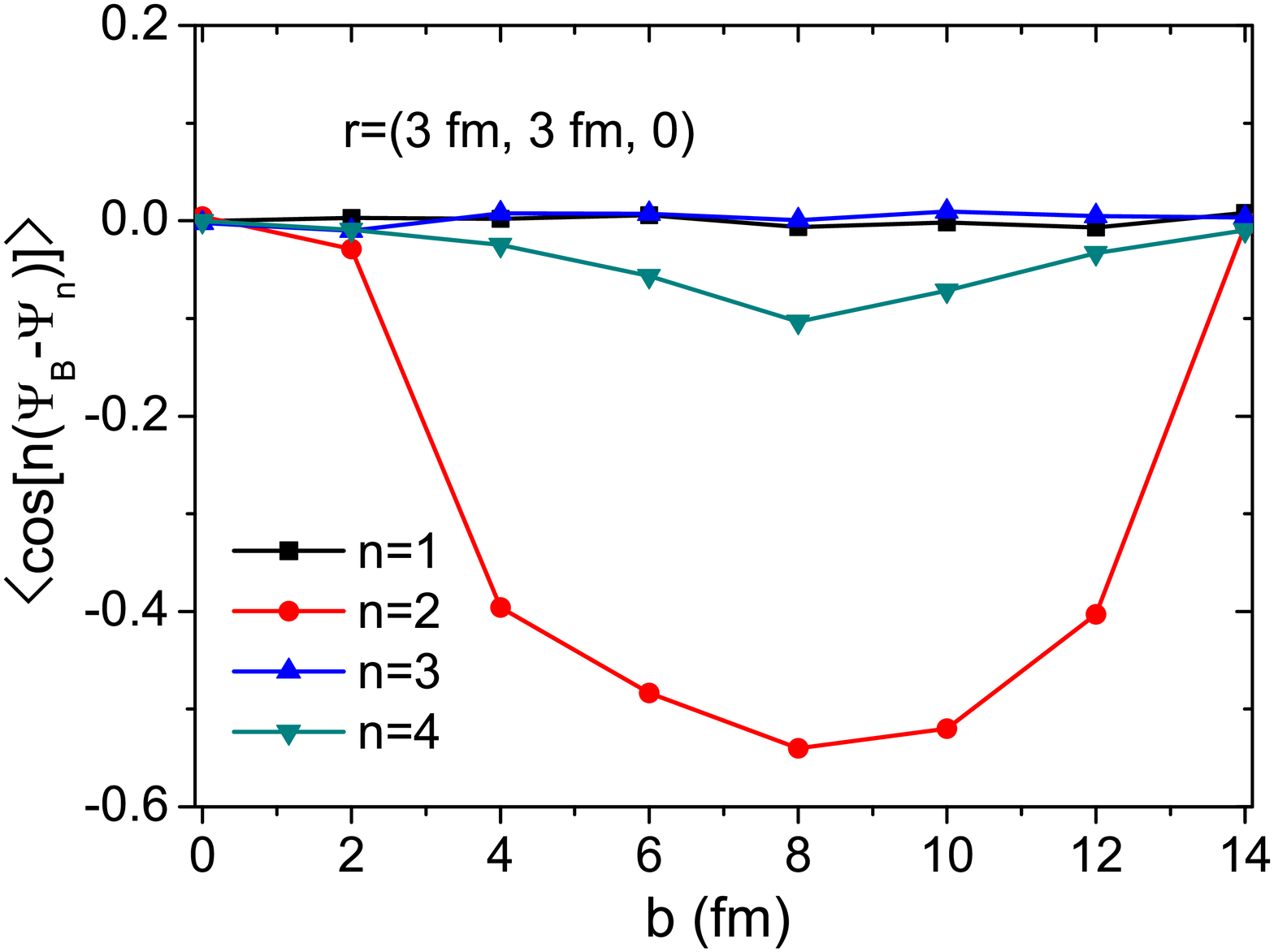}
\caption{(Color online) The correlations $\lan\cos[n(\Psi_{\bf B}-\Psi_n)]\ran$ as functions of impact parameter for $n=1,2,3,4$ at four different positions
on the transverse plane: (from left to right) $\br=(0,0,0)$ fm; $\br=(3,0,0)$ fm; $\br=(0,3,0)$ fm; $\br=(3,3,0)$ fm. }
\label{icos2}
\end{center}
\end{figure*}

\begin{figure}
\begin{center}
\includegraphics[width=5.5cm]{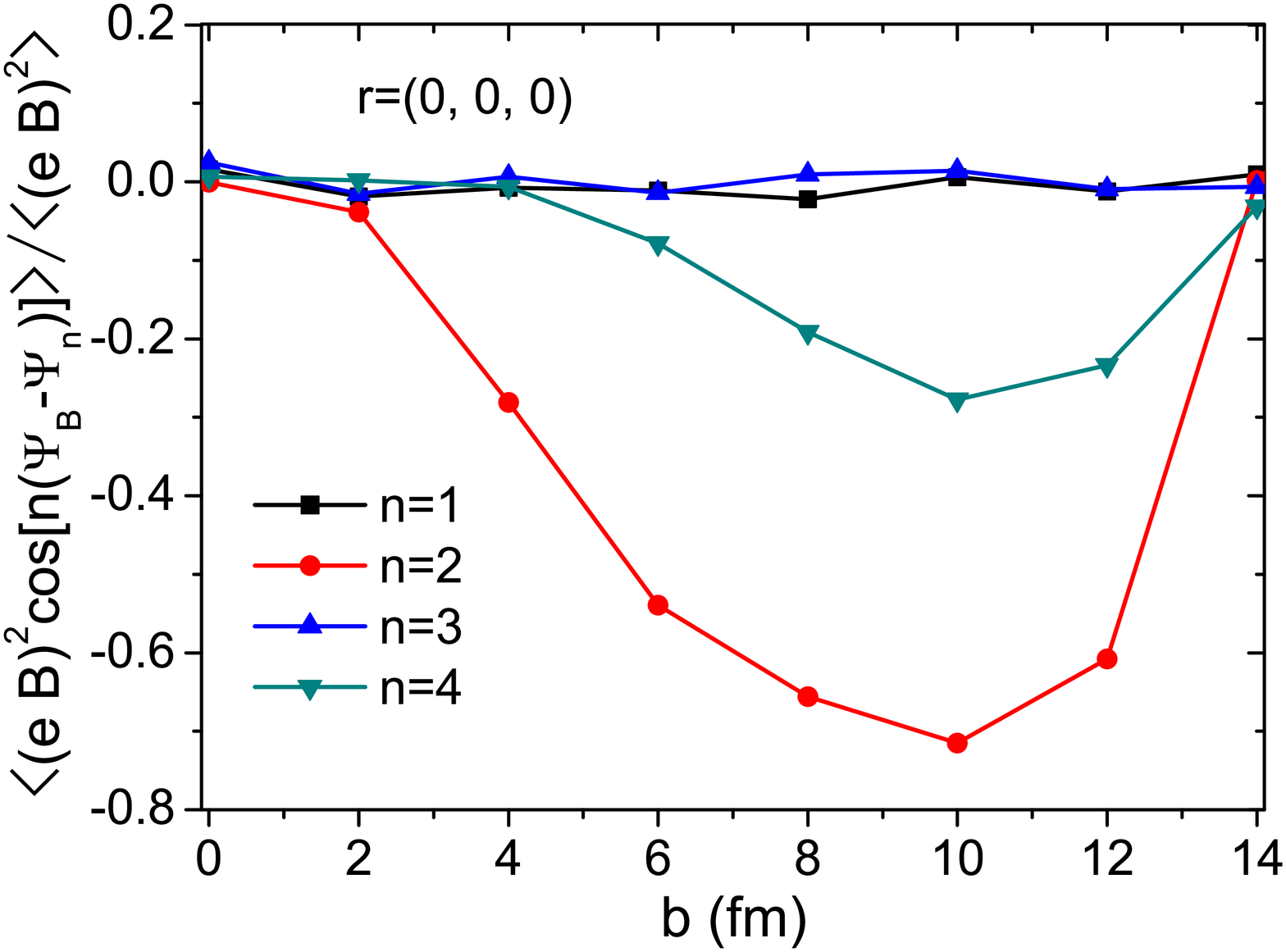}
\caption{(Color online) The   $(e\bB)^2$-weighted
correlations $\lan(e\bB)^2\cos[n(\Psi_{\bf B}-\Psi_n)]\ran/\lan(e\bB)^2\ran$  as functions of impact parameter for $n=1,2,3,4$ at $\br=(0,0,0)$ fm. }
\label{figb2}
\end{center}
\vspace{-0.7cm}
\end{figure}

Recently it has been suggested that a splitting between the elliptic flow of the $\pi^-$ and $\pi^+$ could occur in heavy-ion collisions due to the Chiral Magnetic Wave~\citep{Burnier:2011bf,Burnier:2012ae}, as briefly discussed in Sec.~\ref{intro}. Specifically the CMW will induce an electric quadruple in the net charge distribution along the $\bf B$ field:
\begin{eqnarray}
\r_e(\f) \sim  2r_e \cos[2(\phi - \Psi_{\bf B})] =  2r_e \cos[2(\bar{\phi} - \bar{\Psi}_{\bf B})],
\end{eqnarray}
where $r_e\propto\int d\f\r_e(\f)\cos[2(\phi - \Psi_{\bf B})]$
quantifies the quadruple from CMW. This modifies the final charged hadron distribution by an amount proportional to $r_e$:
\begin{eqnarray}
\frac{dN_\pm(\f)}{d\f}\sim N_\pm\lc 1+ 2v_2\cos(2\bar{\f})\pm\frac{r_e}{N}\cos[2(\bar{\f}-\bar{\J}_\bB)]\rc, \quad
\end{eqnarray}
where $N=N_++N_-$ is the charge multiplicity. Thus one obtains a splitting of the charged pions $v_2$
\begin{eqnarray}
v_2^{\pi^-} - v_2^{\pi^+} = - \, \lan \frac{r_e}{N} \cos(2\bar{\Psi}_{\bf B})\ran.
\end{eqnarray}
Again, we see the modification factor $\sim\lan\cos(2\J_\bB)\ran$ arising from relative orientation between  $\bf B$ and $\J_2$. If $\bar{\Psi}_{\bf B}$ were to be simply $\pi/2$ (i.e. $\bf B$ always out-of-plane) then the sign of splitting is always $v_2^{\pi^-} >  v_2^{\pi^+} $. However due to the fluctuation, $\bar{\Psi}_{\bf B}$ deviates from $\pi/2$ and leads to a smaller splitting.

Finally, we consider the example of recently suggested soft photon production from conformal anomaly in the presence of an external $\bB$ field~\citep{Basar:2012bp}. The photon emitted via this mechanism carries  azimuthal information of the external magnetic field. If one assumes a homogeneous
external $\bf B$ field, then the produced photon spectrum goes like $q_0 dN/d^3{\vec q}\sim (\vec q \times \vec B)^2$ leading to the following  azimuthal distribution:
\begin{eqnarray}
dN_\g/d\phi \propto  \lc 1 - \cos[2(\phi-\Psi_B)] \rc=\lc 1 - \cos[2(\bar{\phi}-\bar{\Psi}_B)]\rc \quad
\end{eqnarray}
which will contribute the following to the elliptic flow of these photons  (with $V_\gamma$ being the signal strength)
\begin{eqnarray}
v_2^\gamma  =   -\lan V_\gamma  \cos (2\bar{\Psi}_{\bf B})\ran.
\end{eqnarray}
Once again we see the same modification factor $\sim \lan \cos(2\bar{\Psi}_{\bf B})\ran$ due to the mismatch between $\bf B$ field and the out-of-plane direction from event to event.  If $\bar{\Psi}_{\bf B}$ were to be simply $\pi/2$ (i.e. $\bf B$ always out-of-plane) then indeed these photons will have positive and sizable elliptic flow. However the fluctuation may bring $\bar{\Psi}_{\bf B}$ to depart from $\pi/2$ considerably.

\begin{figure*}
\begin{center}
\includegraphics[width=4.4cm]{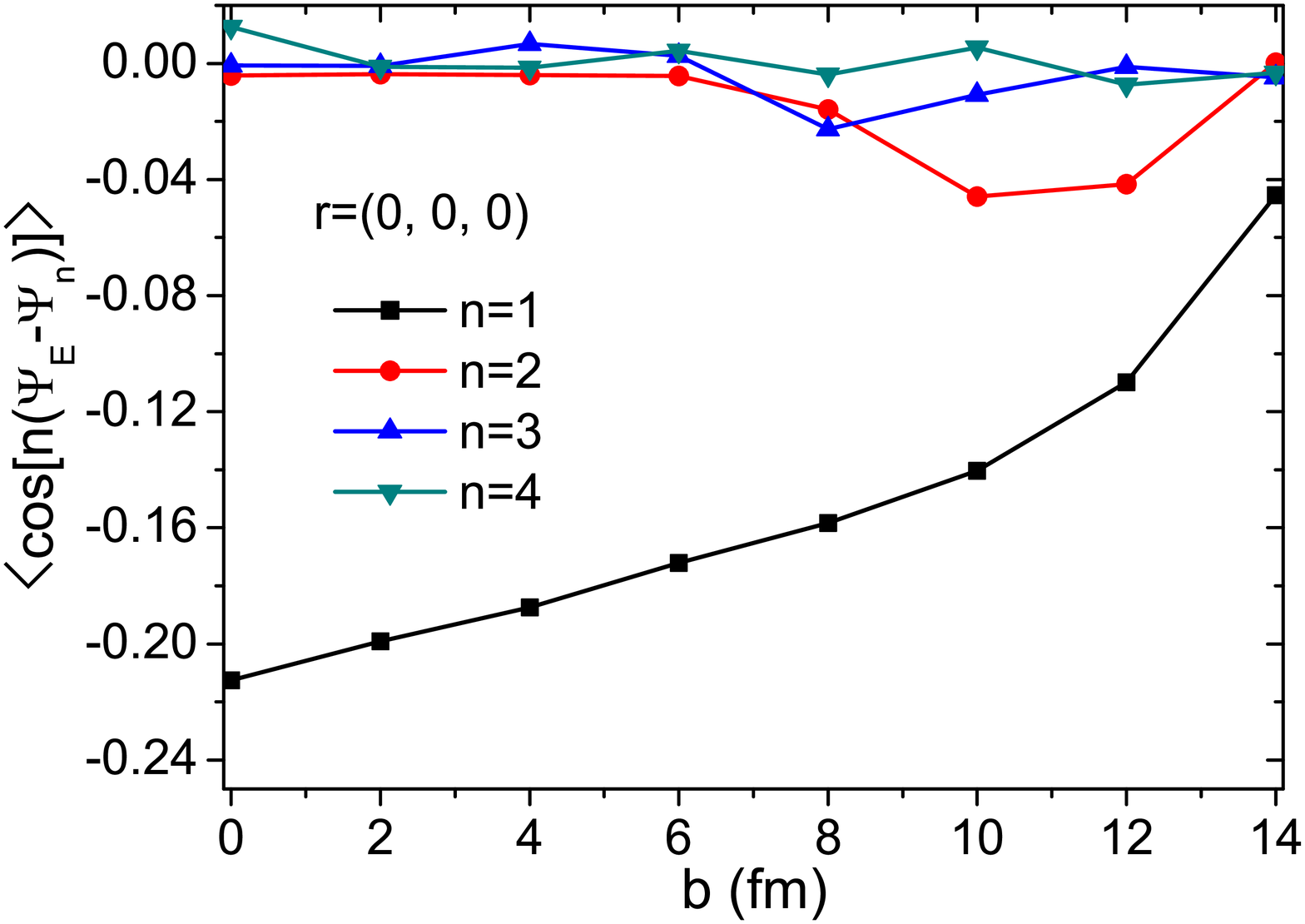}
\includegraphics[width=4.4cm]{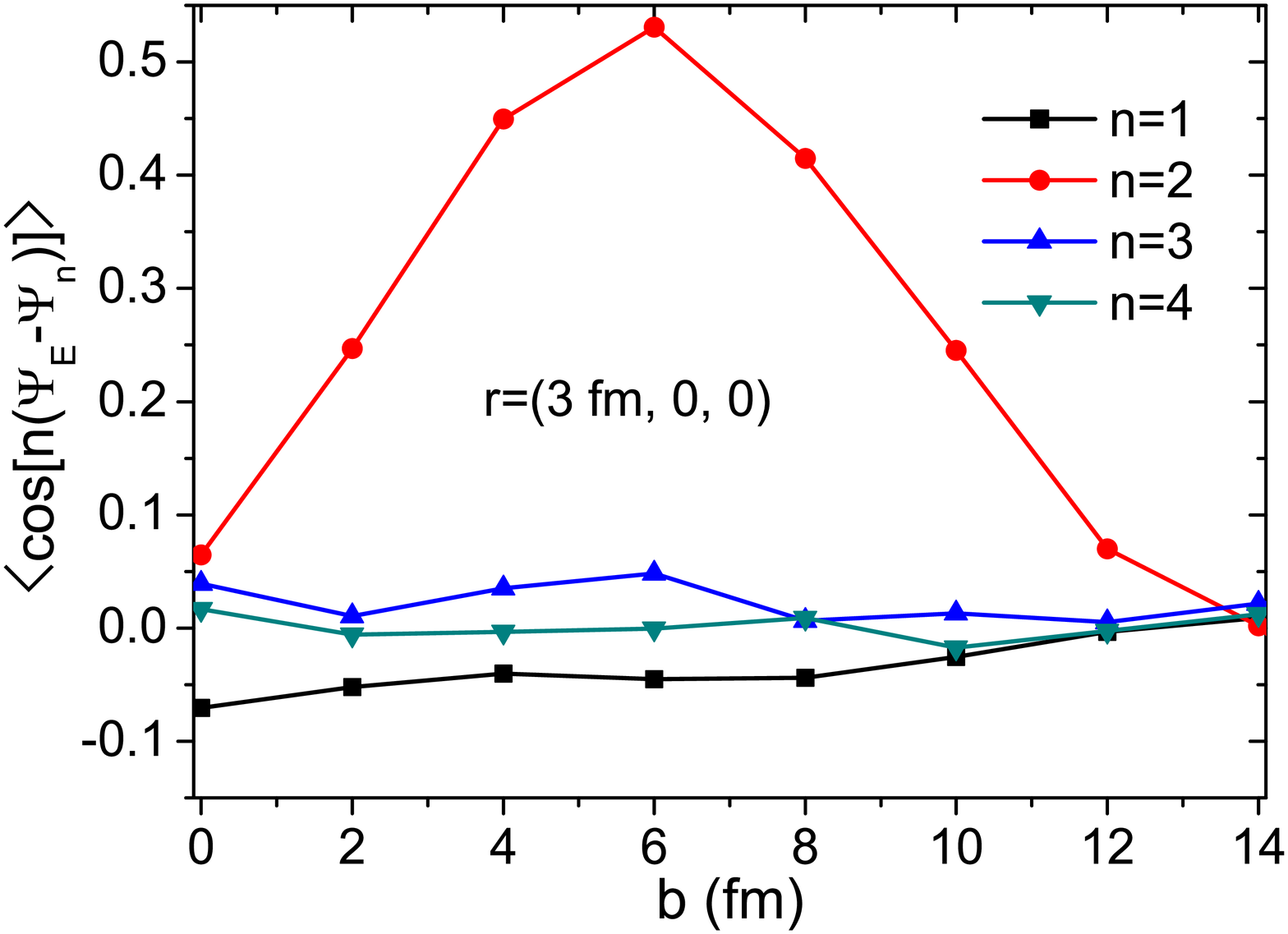}
\includegraphics[width=4.4cm]{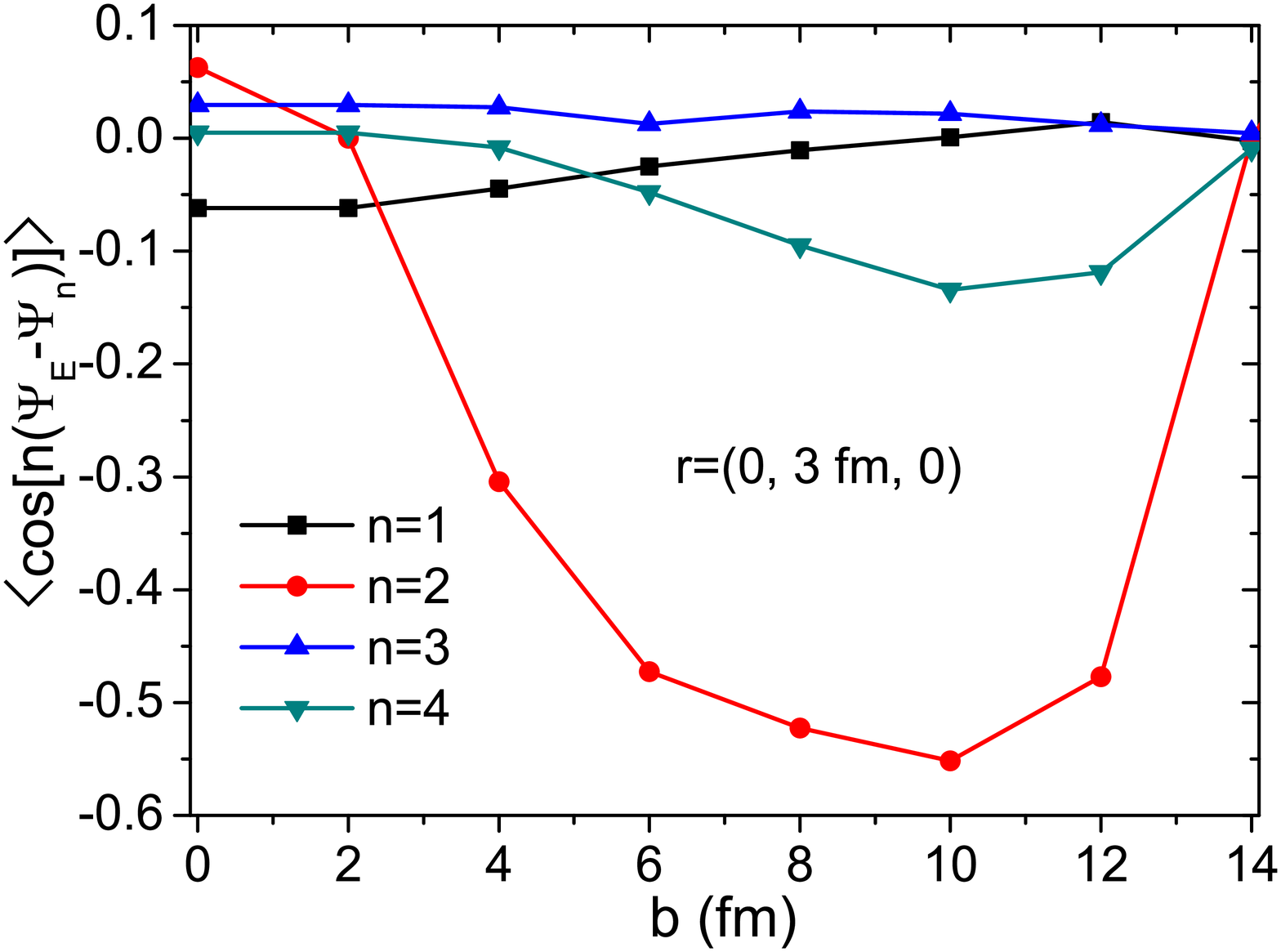}
\includegraphics[width=4.4cm]{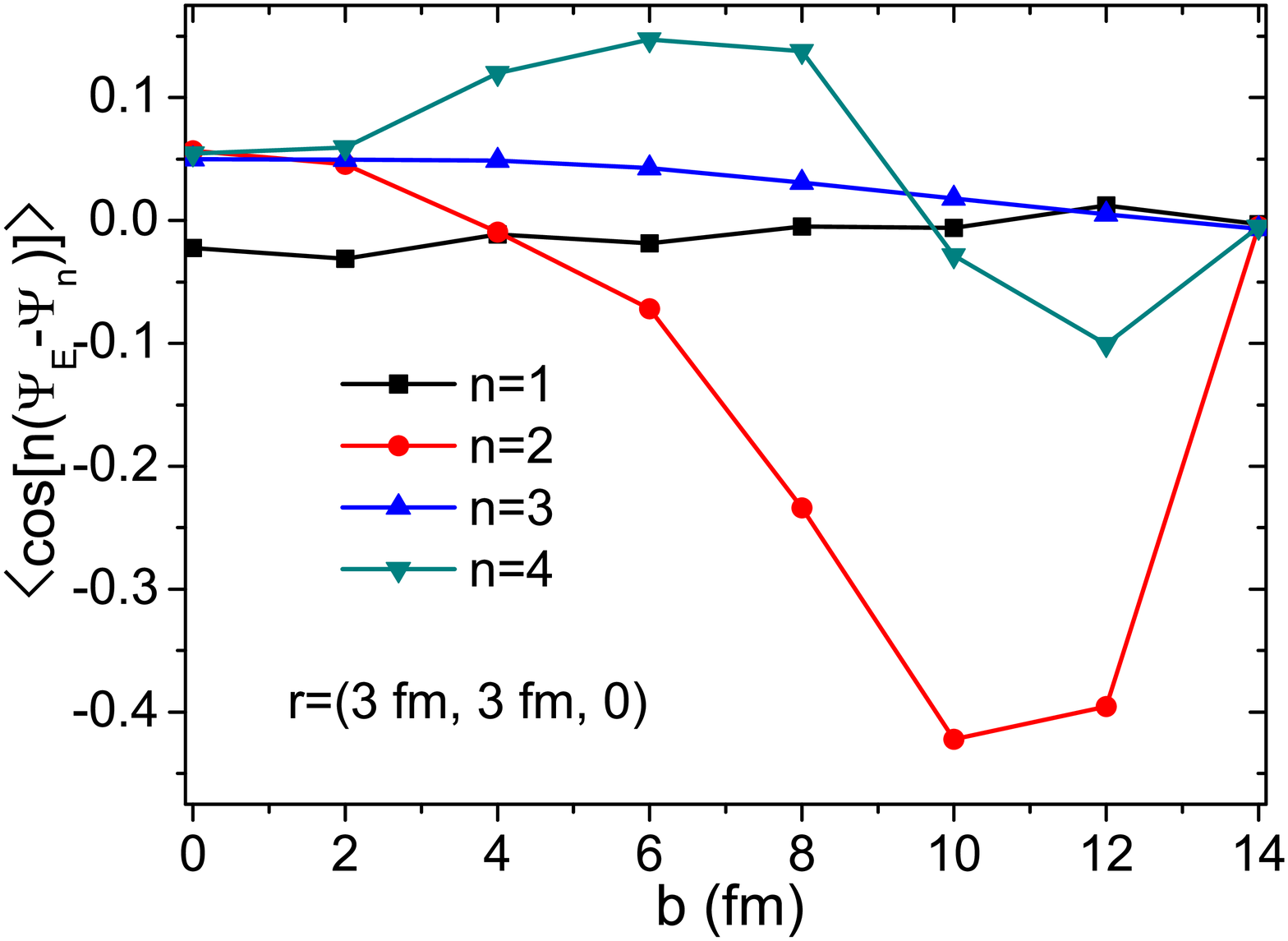}
\caption{(Color online) The correlations $\lan\cos[n(\Psi_{\bf E}-\Psi_n)]\ran$ as functions of impact parameter for $n=1,2,3,4$ at four different positions
on the transverse plane: (from left to right) $\br=(0,0,0)$ fm; $\br=(3,0,0)$ fm; $\br=(0,3,0)$ fm; $\br=(3,3,0)$ fm.  }
\label{fige1}
\end{center}
\end{figure*}

In all three examples, the fluctuating $\J_\bB-\J_2$ brings in a reduction to the intrinsic strength of the signal by
\begin{eqnarray}
\label{reduct1}
R_1=\lan\cos(2\bar{\Psi}_{\bf B})\ran.
\end{eqnarray}
It is therefore worth a close examination of this factor: see Fig.~\ref{icos2}
for the computed average values of  $\lan\cos[n(\Psi_{\bf B}-\Psi_n)]\ran$ for varied centralities from event-by-event determination
of the $\bf B$-field direction $\Psi_{\bf B}$ (at several different spatial points) and the participants harmonics, $\Psi_n$, $n=1,2,3,4$.
The  plots suggest that the correlations  between $\J_\bB$ and the odd harmonics $\J_1, \J_3$ are practically zero (in accord with parity invariance), while
the the correlations of $\J_\bB$ with even harmonics $\J_2, \J_4$ are nonzero but get suppressed as compared with the results in optical geometry limit (without fluctuations) e.g. $\lan\cos[2(\Psi_{\bf B}-\Psi_2)]\ran_{\rm opt}=-1$. The centrality dependence of $\lan\cos[2(\Psi_{\bf B}-\Psi_2)]\ran$ is in perfect agreement with the patterns seen in the histograms Fig.~\ref{his} and scatter plots Fig.~\ref{scat}: it is significantly suppressed in the most central and most peripheral cases (indicating no correlations) while is maximized around $b=8\sim 10$ fm with peak values $-0.6\sim 0.7$.  Similar behavior is observed for $\lan\cos[4(\Psi_{\bf B}-\Psi_4)]\ran$ too albeit with a  weaker correlation strengths, e.g. with the peak values $\sim 0.2$ or so. Although suppressed, the correlation between $\J_\bB$ and $\J_4$ appears still sizable for moderate values of $b$. This may imply complications and caveats that must be seriously addressed for the proposal in Ref.~\cite{Voloshin:2011mx} to use the two-particle correlation with respect to the fourth harmonic event plane to disentangle the CME contribution and the
flow contribution.

The plots in ~Fig.\ref{icos2} for four different field points demonstrate the spatial dependence of the correlations $\lan\cos[n(\Psi_{\bf B}-\Psi_n)]\ran$ on the transverse plane. We first see minor differences when the field point deviates from the origin and such differences become sizable for the outer most point $\br=(3, 3, 0)$ fm. The comparison suggests that further
away from the origin  the correlations between $\J_\bB$ and $\J_n$ become even weaker. However, near
the central overlapping region, the correlations are almost homogeneous over, for example, the distance of the typical size of a sphaleron, $(\a_s T)^{-1}\lesssim 1$ fm for RHIC and LHC~\cite{McLerran:1990de}. 

The last issue we want to address is the possible correlation between the $\bB$ orientation and its own strength. It is conceivable that there could be correlation (on an event-by-event basis) between the signal strength and the $\bf B$ field direction. Typically the signal strength of these $\bf B$-field induced effects will scale as $\sim \bf B^2$. For example, recalling that the CME current is
proportional to $\bB$ and the signal strength $A_{++}$ as mentioned earlier would be nearly proportional to $\bB^2$.
The quadrupole strength $r_e$ through CMW also scales similarly $r_e\propto\bB^2$. So does the signal strength $V_\g$ in the case of soft photon emissions.
If the field strength and the orientation of $\bB$ do correlate with each other, then one may not be able to factorize the signal strength and the field orientation factor $\cos(2\bar{\Psi}_{\bf B})$ when taking the event average. We therefore further examine the correlations between the two by evaluating the factor
\begin{eqnarray}
\label{reduct2}
R_2=\frac{\lan (e\bB)^2\cos(2\bar{\Psi}_{\bf B})\ran}{\lan(e\bB)^2\ran}.
\end{eqnarray}
In Fig.~\ref{figb2} we show the impact parameter dependence of   the $(e\bB)^2$-weighted
correlations $\lan (e\bB)^2\cos[n(\Psi_{\bf B}-\Psi_n)]\ran/\lan(e\bB)^2\ran$ for $n=1,2,3,4$. We find little difference between the two factors $R_1$ in Fig.\ref{icos2}(the most left panel) and $R_2$ in Fig.\ref{figb2} at all centralities for all $n$ values. While only results for field point $\br=(0,0,0)$ are shown in Fig.\ref{figb2}, we have checked all other three points and the observation is the same. Therefore, we conclude that the magnitude of the magnetic field has no noticeable correlation to its azimuthal direction with respect to the matter geometry.

\section {Azimuthal correlation between electric field and matter geometry}\label{elect}
In this section, we briefly discuss how the event-by-event fluctuation affects the correlation between the orientation of $\bE$-field
and the participants harmonic planes, $\J_n$. The motivation is that from Fig.~\ref{field2} we notice that
the electric field can be as strong as the magnetic one, and thus they may lead to observable effects too. An obvious example is possible multiple charge distributions induced by the strong electric field. Similarly to the magnetic field case, possible $\bE$-induced effects would be affected by the azimuthal correlations between $\bE$ and matter geometry.  It is therefore interesting to also study these correlations. In Fig.~\ref{fige1}, we show the correlations $\lan\cos[n(\Psi_{\bf E}-\Psi_n)]\ran$, $n=1,2,3,4$, as functions of the impact parameter at the four field points as before.  There are a number of interesting features that differ from the magnetic field case: 1)  there is a clear negative correlation (i.e. back-to-back) between $\J_\bE$ and $\Psi_1$ and it is most strong in the more central collisions and at the field point near the origin --- this is understandable as the pole of $\Psi_1$ with more matter will concurrently have more positive charges from protons generating $\bE$ pointing in the opposite; 2) at field points away from the origin the $\J_\bE$ is strongly correlated to $\J_2$ (and also weakly correlated to $\J_4$) --- the direction of $\bE$ is more in-plane for the field point on $x$-axis while more out-of-plane for points away from $x$-axis; 3) there is also a weak correlation between $\J_\bE$ and $\J_3$.
Finally we have also checked the $(eE)^2$-weighted correlations
$\lan(e\bE)^2\cos[n(\Psi_{\bf E}-\Psi_n)]\ran/\lan(e\bE)^2\ran$ and find no visible correlation between the $\bE$ magnitude and orientation.

\section{Discussion on charge-dependent measurements}\label{qnc}

The studies of correlations between $\bB$ and $\bE$ fields and matter geometry suggest in general that there may be nontrivial charge distributions, particularly in azimuthal angles (e.g. charged dipole and quadrupole), induced by varied $\bB$- and $\bE-$induced effects. There are also other effects not related the initial $\bE$ and $\bB$ fields, e.g., the local charge conservation effect~\citep{Schlichting:2010qia} that can lead to nontrivial charge azimuthal correlations when coupled with various harmonic flows. It is therefore tempting to think about possible measurements that may fully extract information for the azimuthal charge distributions. In parallel to the measurements of (charge-inclusive) global azimuthal particle distributions that can be subsequently Fourier-decomposed into various harmonic components, here we suggest a class of observables, the charged multiple vectors $\hat{Q}^c_n$.  Consider the measured charged hadrons in an event we can construct the charged multiple vector $\hat{Q}_n^c$ with magnitude $Q^c_n$ and azimuthal angle $\Psi^c_n$:
\begin{eqnarray}
Q^c_n\, e^{in \Psi^c_n} = \sum_i \, q_i \,  e^{i n\phi_i}
\end{eqnarray}
where the summation runs over all particles with $q_i$ and $\phi_i$ the electric charge and azimuthal angle of the $i$-th particle. This idea generalizes the earlier proposal of the charged dipole vector analysis suggested in \citep{Liao:2010nv} in the context of CME observables.
One may introduce properly $p_t$-weighed definition. One may also think of sub-event version of this analysis or possible multi-particle correlation improved version. We emphasize that these are different and independent   from the existing $\hat{Q}_n$ vectors  related to flow measurements, i.e.
$Q_n\, e^{in \Psi_n} = \sum_i \,   e^{in \phi_i}$.
The $\hat{Q}_n$ is charge blind and includes all charges similarly while the $\hat{Q}_n^c$ takes the difference between positive and negative charges therefore yields information on the charge distribution. The technical difficulty of $\hat{Q}_n^c$ measurements should be at the same level as previous $\hat{Q}_n$ analysis and quite feasible, while clearly the $\hat{Q}_n^c$ analysis provides orthogonal and unique information  on the charge distributions. With a joint $\hat{Q}_n$ and $\hat{Q}_n^c$ analysis one can study the strength and azimuthal correlations among all harmonics and charged multipoles. Therefore it is very valuable to do a systematic charged multiple vector analysis, leading toward a quantitative ``charge landscape survey'' in heavy-ion collisions.

\section {Summary}\label{discu}
In summary, we have performed a detailed study of the event-by-event fluctuations of both the azimuthal orientation of
the magnetic and electric fields as well as the matter geometry (which is specified by the participant planes of a series of harmonics) in the initial condition of heavy-ion collisions. Such fluctuations suppress
the azimuthal correlations between the magnetic field and the second harmonic participant plane particularly in the very central and very peripheral collisions, while leaving a window around $b=8\sim 10$ fm with  still
sizable correlation between the two. We have further studied similar azimuthal correlations between $\bB$ and other harmonic participant planes, and found similar yet weaker correlation with $\J_4$ while no correlation with $\J_1$ and $\J_3$. The correlation between $\J_\bB$ and $\J_4$ may indicate that the CME can also contribute to the charge-pair correlation $\lan\cos[2(\phi_1+\phi_2-2\Psi_{4})]\ran$ as opposed to the assumption in Ref.~\cite{Voloshin:2011mx}.
Examination of these correlations at different field points shows notable dependence on spatial positions. For completeness we have presented similar studies for electric field $\bE$ azimuthal correlations with matter geometry which show quite different patterns from that for the $\bB$ field.
We have evaluated the impact of  such azimuthal fluctuations and correlations on a number of observables related to magnetic-field induced effects in heavy-ion collisions: the charged pair azimuthal correlations,
the charge dependent elliptic flow of pions, as well as the azimuthal anisotropy of soft photons due to conformal anomaly.
Specifically we have quantified
the modification factors $R_1=\lan\cos(2\bar{\Psi}_{\bf B})\ran$ and the $(e\bB)^2$-weighted reduction factor $R_2=\lan(e\bB)^2\cos(2\bar{\Psi}_{\bf B})\ran/\lan(e\bB)^2\ran$ at
different centralities (see Figs.~\ref{icos2} and \ref{figb2}) and found sizable reduction in both very central and very peripheral collisions. From these results we conclude that
 the optimal centrality class for observing the above mentioned $\bB$-induced effects corresponds to impact parameter range $b\sim8-10$ fm.
The qualitative conclusion should hold also for Pb + Pb collisions at LHC.

We end with a few pertinent remarks:\\
(1) In our computations (as well as all the previous computations~\cite{arXiv:0711.0950,arXiv:0907.1396,arXiv:1003.2436,arXiv:1103.4239,arXiv:1107.3192,arXiv:1111.1949,arXiv:1201.5108}),
 the classical Li\'enard-Wiechert potentials have been used. One may worry about quantum corrections to the field equations as the magnitude of the electromagnetic field
is much larger than the electron and light quark masses. In principle a calculation including all relevant QED processes is needed, but a one-loop Euler-Heisenberg effective lagrangian (see Ref.~\cite{Dunne:2004nc} for review) may give a good indication on the size of such corrections. At strong-field limit the field equation derived from this lagrangian can be regarded as linear Maxwell equations with the renormalized charge $e^2\rightarrow \tilde{e}^2=e^2/\ls1-\frac{e^2}{24\p^2}\ln\frac{e^2|F|^2}{m_e^4}\rs$. An estimate implies that the computed electromagnetic field may be amended only by a few
 percent even for $e|F|\sim 100 m_\p^2$.\\
(2) As shown in Refs.~\cite{arXiv:1103.4239,arXiv:1201.5108}, the magnetic field magnitude bears strong time dependence. Furthermore, the electromagnetic response of QGP could dramatically modify the time evolution
of the $\bB$ and $\bE$ fields. It would be important to incorporate these into realistic modeling of effects induced by these fields, as recently attempted in ~\cite{Toneev:2012zx}   to incorporate the time evolution into
the event-by-event estimates of the CME by using the parton-hadron-string-dynamics approach.\\
(3) As suggested in Ref.~\cite{Voloshin:2010ut}, in order to disentangle the effects of the magnetic
field and the collective flow, it is very useful to study the U + U collisions where the geometry becomes a nontrivial level arm. It will be interesting to study in the future the various correlations among the field strengths, field azimuthal orientations with the matter geometric eccentricities as well as the harmonic participant planes.

{\bf Acknowledgments:}
We are grateful to Y. Burnier, A. Bzdak, U. Heinz, D. Kharzeev, V. Koch, V. Skokov, H. Yee, and Z. Qiu for helpful communications and discussions. We also thank the RIKEN BNL Rsearch Center for partial support.
\vskip0.3cm


\end{document}